# On-chip cavity electro-acoustics using lithium niobate phononic-crystal resonators

Jun Ji[1], Joseph G. Thomas[1], Zichen Xi[1], Liyang Jin[1], Dayrl P. Briggs[2], Ivan I. Kravchenko[2], Arya G. Pour[3], Liyan Zhu[1,4], Yizheng Zhu[1], and Linbo Shao[1,3,*]

[1] *Bradley Department of Electric and Computer Engineering, Virginia Tech, Blacksburg, VA, USA*
[2] *Center for Nanophase Materials Sciences, Oak Ridge National Laboratory, Oak Ridge, TN, USA*
[3] *Department of Physics and Center for Quantum Information Science and Engineering (VTQ), Virginia Tech, Blacksburg, VA, USA*
[4] *Center for Power Electronics Systems (CPES), Virginia Tech, Blacksburg, VA, USA.*
* Email: shaolb@vt.edu

## Abstract

Mechanical systems are pivotal in quantum technologies because of their long coherent time and versatile coupling to qubit systems. So far, the coherent and dynamic control of gigahertz-frequency mechanical modes mostly relies on optomechanical coupling and piezoelectric coupling to superconducting qubits. Here, we demonstrate on-chip cavity electro-acoustic dynamics using our microwave-frequency electrically-modulated phononic-crystal (PnC) resonators on lithium niobate ($LiNbO_3$, LN). Leveraging the high dispersion of PnC, our acoustic modes space unevenly in the frequency spectrum, emulating atomic energy levels. Atomic-like transitions between different acoustic modes are selectively achieved by applying electrical fields to modulate acoustic modes via nonlinear piezoelectricity of LN. Among two modes, we demonstrate Autler–Townes splitting (ATS), alternating current (a.c.) Stark shift, and Rabi oscillation with a maximum cooperativity of 4.18. Extending to three modes, we achieve non-reciprocal frequency conversions with an isolation up to 20 dB. Nonreciprocity can be tuned by the time delay between the two modulating pulses. Our cavity electro-acoustic platform could find broad applications in sensing, microwave signal processing, acoustic computing, and quantum acoustics.

***Introduction*** – Microwave-frequency acoustic waves in solids are featured with long lifetimes [1], coherent times [2], small footprint (micron scale at gigahertz frequencies, compared to centimeter scale of their electromagnetic (EM) counterparts), and coupling to different systems such as photonics [3-5], magnons [6,7], superconducting qubits [8-12], and solid-state electronic spins [13-15]. Benefitting from these features, emerging integrated acoustic-wave devices and circuits enable applications such as on-chip quantum networking [16-18], phononic quantum memories [19,20], microwave-to-optical conversions [21-24], low-phase-noise oscillation [25,26], magnetic-free isolation and circulation [27-31], acousto-optic lidar [32], and analog computing [33,34].

While high-$Q$ microwave-frequency acoustic resonant modes [25,35-37] are widely utilized in these applications, dynamic, direct, and selective interactions between different modes are still challenging but desired, for example, in building phononic quantum gates. Previous demonstrations of such interactions have relied on mediation by other systems, such as parametrically driven superconducting qubits [38]. Opto-mechanical cavities [39,40] and electro-mechanical cavities [41,42] have also provided dynamic interactions between acoustic modes, while these modes are limited to megahertz frequencies. Electro-acoustic modulation, leveraging nonlinear material properties of LN [43], offers a dynamic and direct control of microwave-frequency travelling acoustic waves [44,45].

In this Letter, we demonstrate an on-chip cavity electro-acoustic platform, providing dynamic, direct, and selective control of acoustic waves among discrete atomic-like energy levels. The 1-GHz PnC modes are spectrally resolved and unevenly spaced, featuring high quality ($Q$) factors up to 12,000. Leveraging the electro-acoustic effect of LN, atomic-like transitions between two neighboring modes are directly and selectively achieved. Electrically modulating the modes, we experimentally observe ATS, a.c. Stark effects, and Rabi oscillations. We achieve a strong coupling between two acoustic modes, showing a maximum cooperativity of 4.18. Extending to three-level acoustic systems, we demonstrate programmable non-reciprocal frequency conversions with an isolation up to 20 dB by sequentially applying two π pulses.

***Electrically modulated PnC resonators –*** Our integrated electro-acoustic devices are fabricated on an *X*-cut LN substrate with a patterned silicon nitride (SiN) thin film on top (**Fig. 1(a)**). The SiN pillars form a one-dimensional (1D) PnC resonator in an acoustic waveguide structure. Details of device designs and characterization are discussed **in Sec. 1** and **Fig. S1** of **Supplemental Material**. Leveraging the strong dispersion at the upper edge of the PnC bandgap, our PnC resonator supports multiple modes [25,35-37] that are spectrally unevenly spaced (**Fig. 1(b)** and **1(c)**). Thus, transitions between two specific modes can be selectively driven by electro-acoustic modulation frequency. In comparison, a single acoustic ring resonator [46,47] typically support modes with a nearly constant free spectral range (FSR), which complicates transition-selective driving. Alternatively, a nonuniformly-spaced *N*-level system can be formed by *N* coupled ring resonators [48], which, however, complicates the design and fabrication as *N* increases.

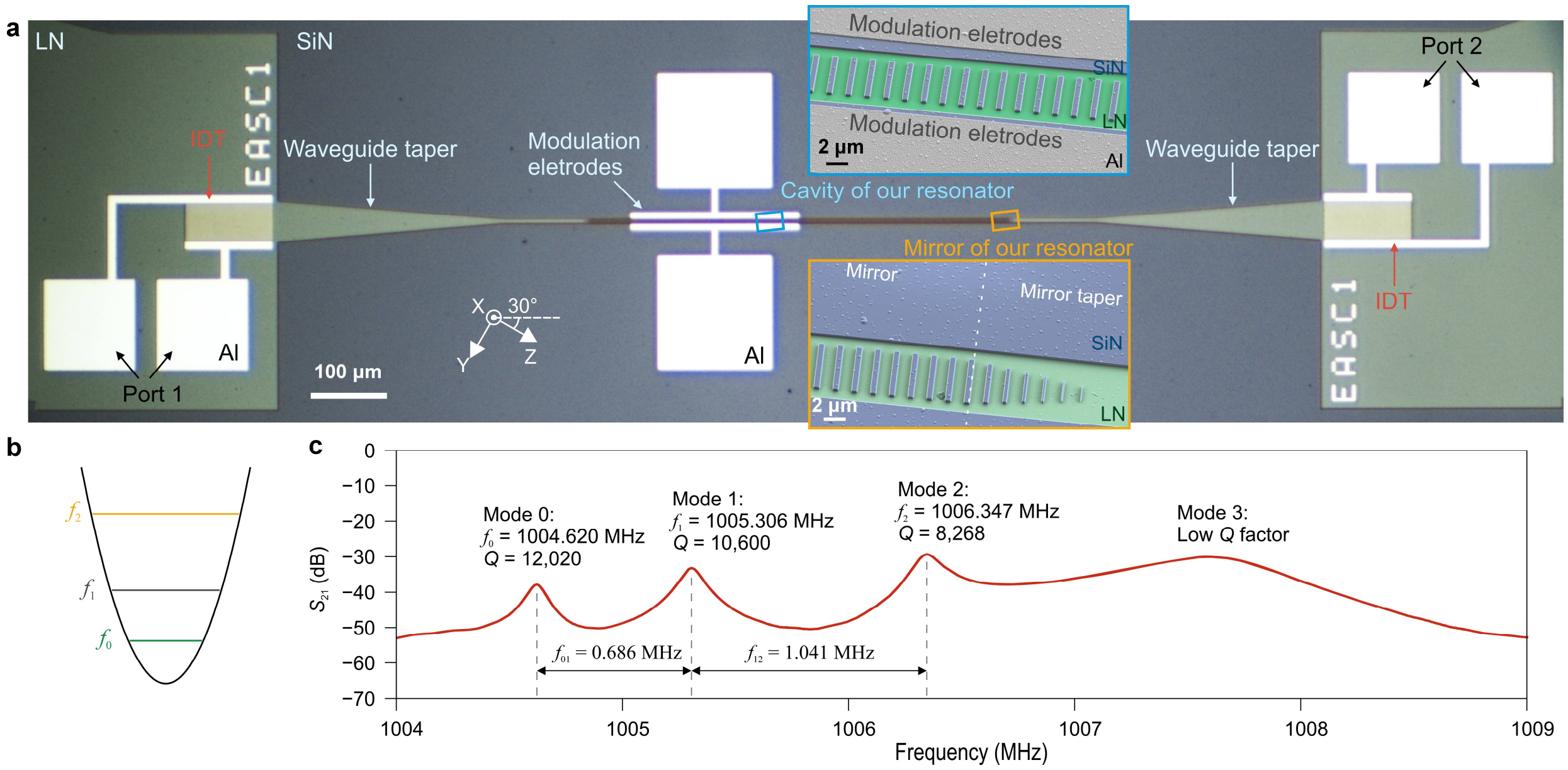

**FIG. 1. The on-chip cavity electro-acoustic platform.** (**a**) Optical micrograph of the fabricated phononic-crystal (PnC) resonator on an acoustic waveguide. Inset: False colored scanning electron microscopy (SEM) images of different regions. The device is oriented 30° with respect to the crystal Z-axis. (**b**) Our PnC resonator supports three high-*Q* modes that are unevenly distributed in frequency. (**b**) Measured transmission spectrum $S_{21}$ of our PnC resonator.

Our PnC resonator supports three modes (Mode 0, 1, and 2) at $f_0$ = 1,004.620 MHz, $f_1$ = 1,005.306 MHz, and $f_2$ = 1,006.347 MHz (**Fig. 1(c)**). The spectral spacing between Mode 0 and Mode 1 ($f_{01}$ = 0.686 MHz) differs from that between Mode 1 and Mode 2 ($f_{12}$ = 1.041 MHz). The loaded *Q* factors of these modes are 12,020, 10,600, and 8,268; time domain ring-down measurements **(Fig. S2)** show long lifetimes up to 1.966 μs, which agrees with the spectrally measured *Q* factors. The difference between the spectral spacing $f_{12}$ - $f_{01}$ is 0.355 MHz, which is significantly larger than any linewidth of the modes, for example, linewidth of Mode 0 is $f_0/Q$ = 0.08358 MHz. By tuning the cavity length, both the number of spectrally resolved high-*Q* modes and their uneven frequency spacings can be engineered (**Fig. S3**).

The modulation electrodes cover only half of the resonator, providing an (equivalently) anti-symmetric modulation via the electro-acoustic effect [44,45] (**Sec. 8** of **Supplemental Material**). Given the symmetric design of our PhC resonator, the fundamental mode (Mode 0) and second-order higher mode (Mode 2) exhibit symmetric strain profiles, whereas the first-order higher mode (Mode 1) is anti-symmetric. Their displacement profiles are measured using our in-house microwave-frequency optical vibrometer [49] (**Fig. S1(f)-(h)**). In this configuration, the symmetry-based selection rule allows interactions between neighboring modes (i.e., between Mode 0 and Mode 1, between Mode 1 and Mode 2) but forbids transition between next-neighboring modes (i.e., between Mode 0 and Mode 2).

Our electrically-modulated PnC resonator is analogous to a driven atomic system. The spectrally unevenly-spaced acoustic modes play the role of anharmonic energy levels. The electro-acoustic modulation coherently couples neighboring modes, analogous to transitions between energy levels. When the modulation frequency matches (or is slightly detuned from) the frequency spacing between modes, the system corresponds to on-resonance (or detuned) driving. The intensities in each acoustic mode are analogous to the population occupying the atomic energy levels. Within this analogy, we experimentally demonstrate acoustic-wave counterparts of ATS, a.c. Stark shifts, and Rabi oscillations. We note that this work uses coherent sources to pump the acoustic modes, simulating the classical limit of a cluster of atoms.

***Spectral dynamics of electrically modulated acoustic modes*** – We demonstrate the direct control between two modes (Mode 0 and Mode 1) using the electro-acoustic modulation. Acoustic ATS, by which the two acoustic energy levels split into four, is observed in spectra near resonant frequencies of Mode 0 and Mode 1 when the electrical modulation frequency $f_m$ matches the frequency spacing $f_{01}$ between them (**Fig. 2(a)**). Acoustic ATS becomes larger as the modulation amplitude increases. The difference in the heights of split transmission peaks can be attributed to the slight difference between the applied modulating frequency $f_m = 701$ kHz and the actual frequency gap between two modes $f_{01} = 686$ kHz. Meanwhile, the applied modulating signal of $f_m = 701$ kHz is largely detuned from the transition frequency $f_{12} = 1.041$ MHz between Mode 1 and Mode 2, thus no significant changes in the transmission spectra are observed near the resonant frequency of Mode 2. To illustrate the mode mixing by the electrical modulation, we characterize out-of-plane displacement profile when our PnC resonator is pumped continuously at $f_0$ and modulated continuously at $f_{01}$, showing both Mode 0 and Mode 1 are excited (**Movie S1**).

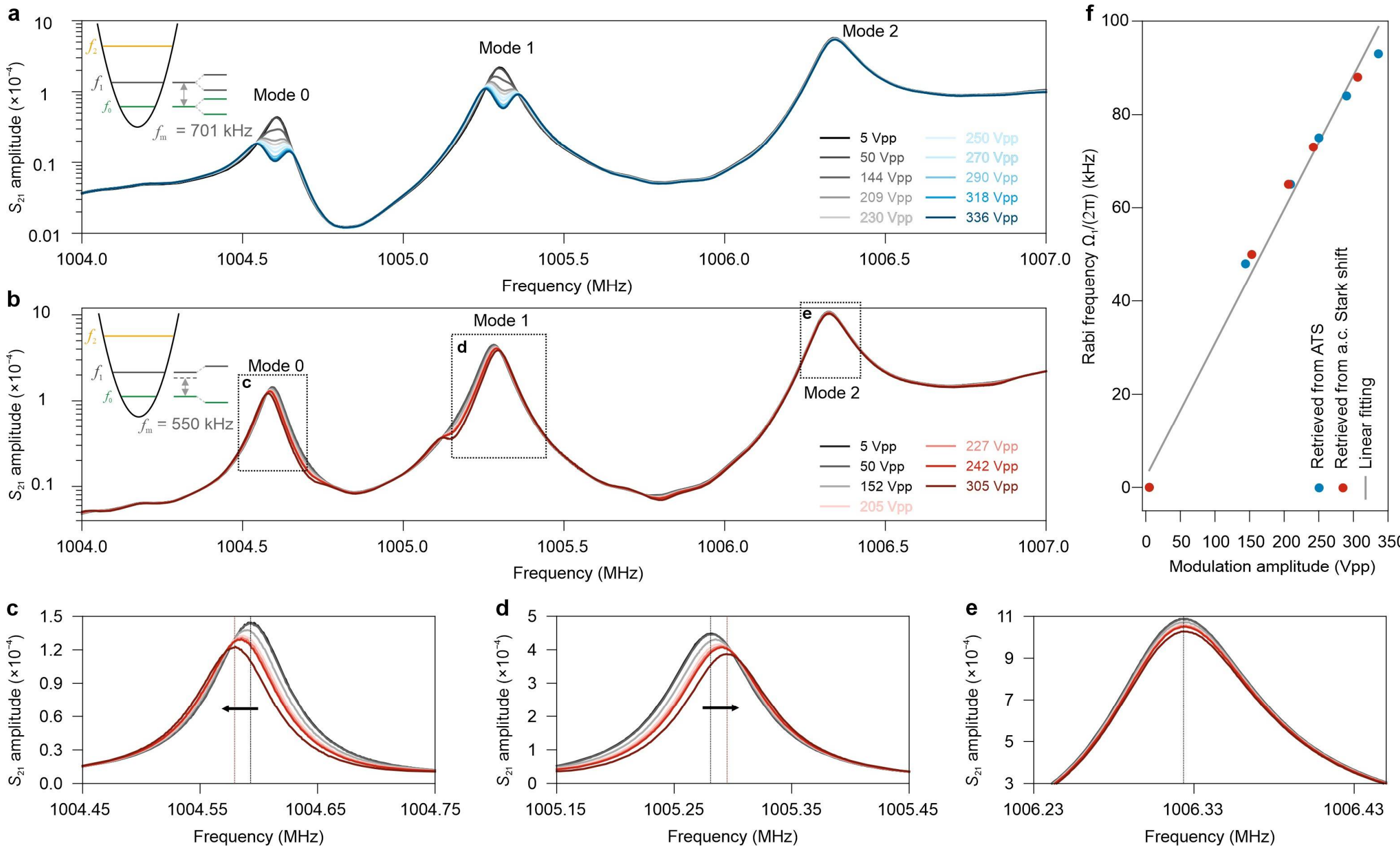

**FIG. 2. Spectral dynamics of electrically modulated acoustic modes.** (**a**) When the modulation frequency $f_m$ closely matches the frequency separation $f_{01}$, acoustic ATS is observed. Here, $f_m = 701$ kHz and $f_{01} = 686$ kHz. (**b**) When the modulation frequency $f_m$ is far detuned from $f_{01}$, acoustic a.c. Stark shifts are observed. Here, $f_m = 550$ kHz and $f_{01} = 686$ kHz. (**c**) to (**e**) Closed-up views show that Mode 0 and Mode 1 experience a.c. Stark shifts, while Mode 2 does not experience a.c. Stark shifts. (**f**) Rabi frequencies extracted from acoustic ATS and ac. Stark shifts measurements.

Acoustic a.c. Stark shifts are observed in the transmission spectra when the electrical modulation of $f_m$ = 550 kHz is red-detuned from the transition frequency $f_{01}$ = 686 kHz (**Fig. 2(b)**). With the increasing modulating amplitude, the peak transmission frequencies related to Mode 0 and Mode 1 shift apart from each other (**Fig. 2(c)** and **2(d)**). At a modulating amplitude of 305 peak-to-peak voltage (Vpp), we observed a maximum red shift of 13 kHz for Mode 0 and blue shift of 14 kHz for Mode 1. Meanwhile, no shift in the resonant frequency of Mode 2 is observed (**Fig. 2(e)**) as it is largely detuned from any transitions. Similar results are experimentally observed between Mode 1 and Mode 2. (**Fig. S5**). On the other hand, direct transition between Mode 0 and Mode 2 is prohibited by the selection rule (**Fig. S6**). The dynamics of our system are further discussed in **Sec. 5** of **Supplemental Material**. The experimental results agree with our analytical model (**Fig. S4**). Rabi frequencies extracted from ATS and a.c. Stark shifts show consistent and expected linear dependency on applied modulating amplitudes (**Fig. 2(f)**).

***Coherent temporal dynamics of two acoustic modes* –** We demonstrate coherent and dynamic control of our acoustic system by performing Rabi oscillation between two acoustic modes (**Fig. 3(a)**). We pump our PnC resonator at the resonant frequency $f_0$ of Mode 0 for 150 μs. We then apply an electrical modulation with its frequency $f_m$ matching the transition frequency $f_{01}$ for 280 μs (**Fig. 3(b)**). Meanwhile, we continuously monitor the output of the PnC resonator at Port 2 (**Fig. 1(a)**); the amplitude of each mode is extracted by bandpass filtering the output signal near resonant frequency of each mode (see **Sec. 2 of Supplemental Material**).

We observe Rabi oscillations between two modes (**Fig. 3(c)**). At a modulation amplitude of 250 Vpp, the temporal period of the energy exchange between Mode 0 and Mode 1 is 13.9 μs, which corresponds to Rabi frequency $\Omega_1/2\pi$ of 72 kHz. When the modulation amplitude increases from 250 to 337 Vpp, Rabi frequency $\Omega_1/2\pi$ increases linearly from 72 kHz to 91 kHz (**Fig. 3(d)**). These results agree with parameters extracted from the ATS and a.c Stark shift measurements shown in **Fig. 2(f)**. The measured temporal dynamics agree with the simulations based on the analytical model. We note that Rabi frequency achieved here is significantly larger than that (typically only a few kHz) of parametric coupling in MHz-frequency mechanical systems [41,42].

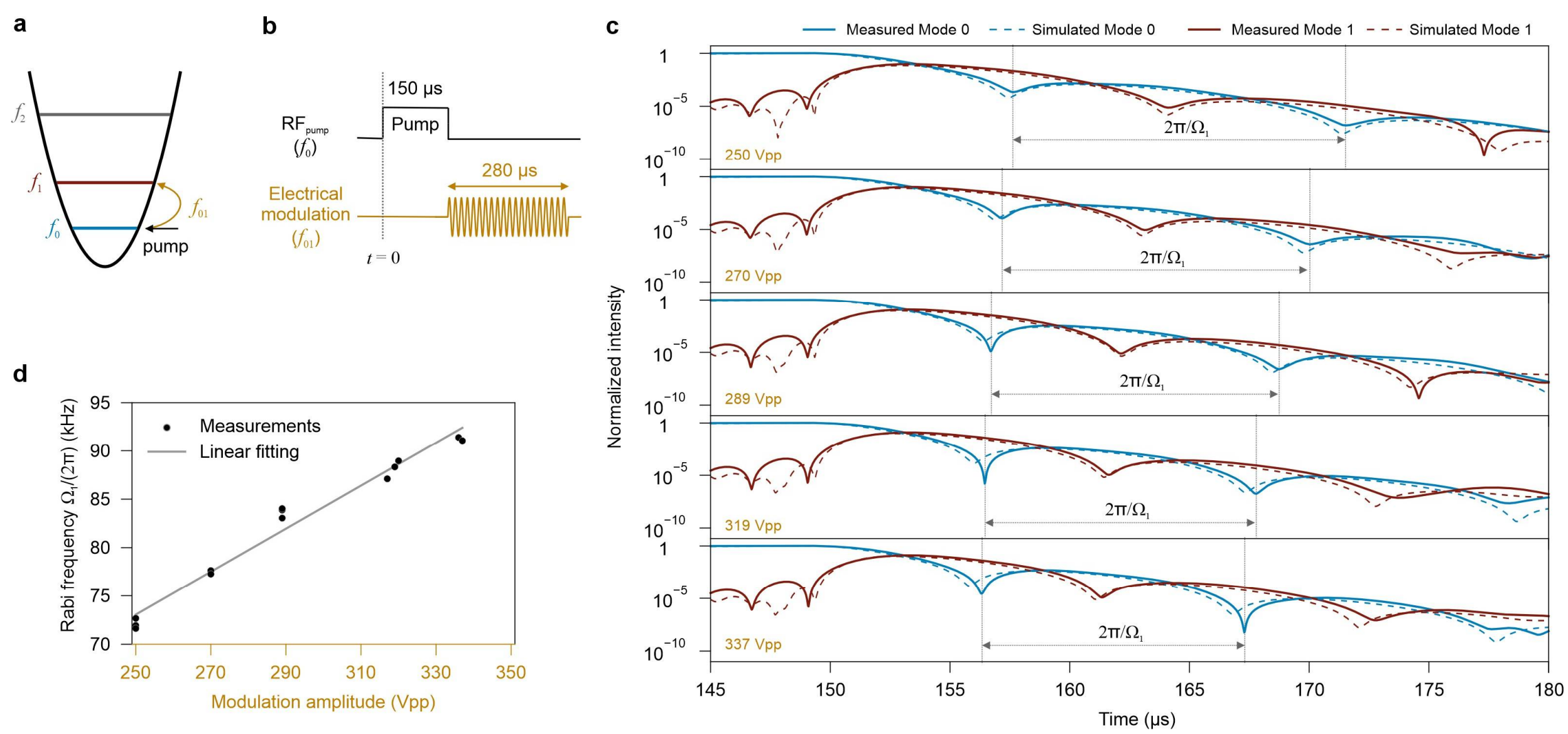

**FIG. 3. Coherent temporal dynamics of two acoustic modes.** (**a**) Rabi oscillation between Mode 0 and Mode 1, when Mode 0 is pumped and modulated by an electrical pulse with its frequency $f_m$ close to $f_{01}$. (**b**) Pulse sequence used for Rabi oscillation measurements. (**c**) One-shot measurements and numerical simulations of the normalized intensity at Mode 0 and Mode 1, as the amplitude of the electrical modulating pulse varies. The electrical modulation turns on at $t$ = 150 μs. The temporal distance between the first two intensity dips relates to the Rabi frequency as $2\pi/\Omega_1$. (**d**) Extracted Rabi frequencies $\Omega_1/(2\pi)$ at different modulation amplitude, with $f_0$ =1004.68 MHz and $f_m$ = 0.699 MHz.

We achieve electro-acoustically-induced strong coupling between two microwave-frequency mechanical modes with a maximum cooperativity, defined as $C \equiv 4\Omega_1^2/(\gamma_0\gamma_1)$, of 4.18 (**Table. S1**). By further increasing the cavity length of the resonator (**Fig. S3**), our PnC resonator shows a smaller frequency difference between modes (e.g., 432 kHz) while maintaining high $Q$ factors (e.g., 14,907), which gives access to a stronger coupling regime where Rabi frequencies are close to or even exceed the transition frequency, i.e., $\Omega_1/2\pi \sim f_{01}$. Thus, our acoustic platform could be used to explore the ultra-strong coupling region, where the rotating-wave approximation is no longer valid [50].

***Programmable non-reciprocal frequency conversions*** – Extending to a three-level system supported by our PnC resonator, we demonstrate programmable non-reciprocal frequency conversions between Mode 0 and Mode 2, assisted by Mode 1 (**Fig. 4(a)**). This is achieved by electrically applying a pair of π pulses. First pulse is of a carrier frequency $f_{01}$ with total pulse duration $t_{01}$ and the second pulse at $f_{12}$ with duration $t_{12.}$ The time between the peaks of two pulses defines the delay time $t_{\text{delay}}$, for example, $t_{\text{delay}}$=0 indicates two modulation pulses are completely overlapped.

We consider a specific case that the $f_{12}$ pulse is applied immediately after the $f_{01}$ pulse, i.e. $t_{\text{delay}} \sim (t_{01} + t_{01})/2$. When the input signal is at the resonant frequency of Mode 0, Mode 0 is excited and acoustic waves in Mode 0 will be sequentially and efficiently transited to Mode 1 and Mode 2 by the modulating pulses (left in **Fig. 4(a)**). In contrast, when the input signal is at the resonant frequency of Mode 2, Mode 2 is excited but acoustic waves in Mode 2 will not be efficiently transited to Mode 0 by the modulating pulses due to the mismatched modulating frequencies with transition frequencies (right in **Fig. 4(a)**).

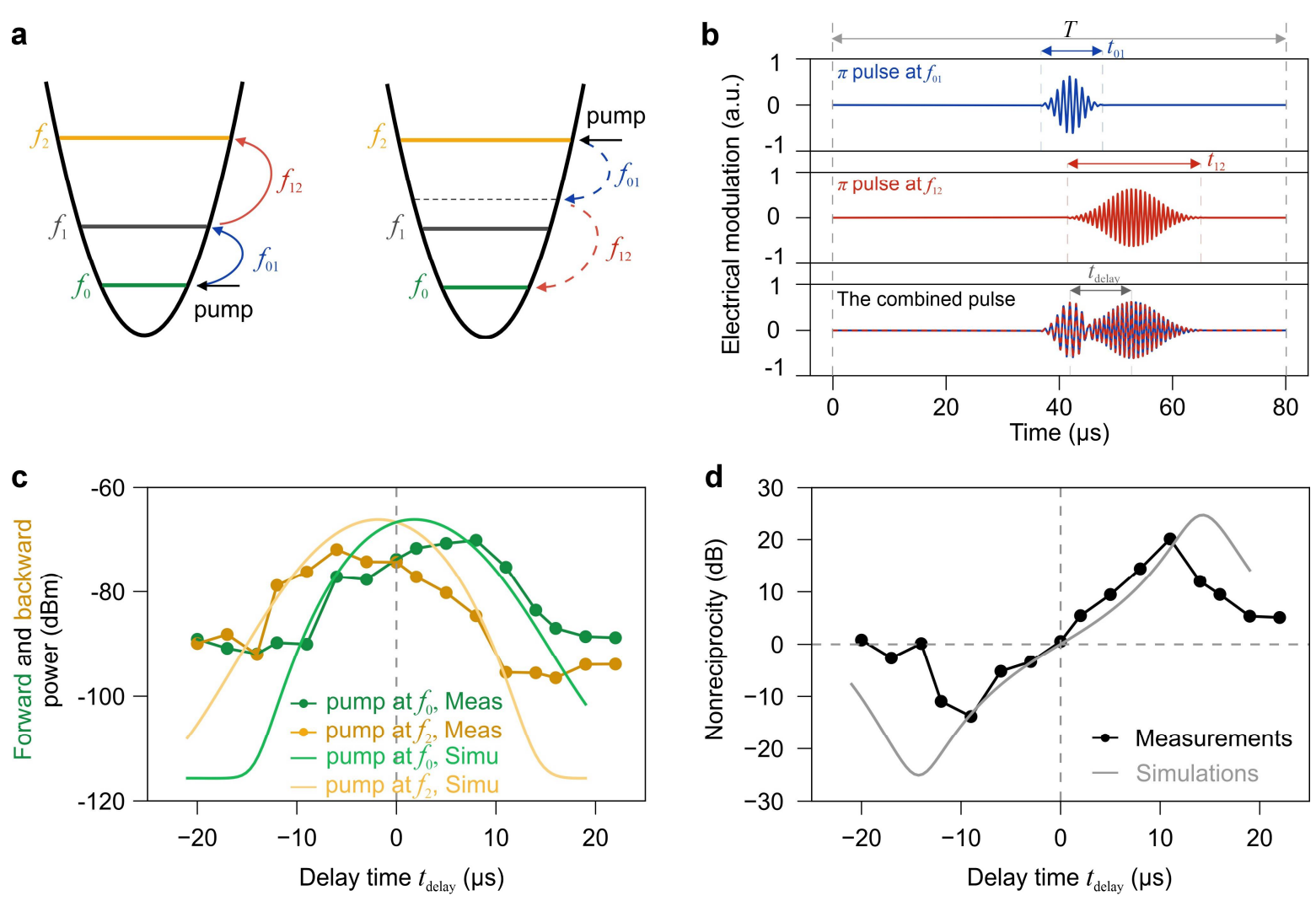

**FIG. 4. Programmable non-reciprocal frequency conversion using three acoustic modes.** (**a**) Diagram showing the non-reciprocal frequency conversion. (**b**) The pulse sequence repeated at $1/T$ for non-reciprocal frequency conversions: One π pulse is of a carrier frequency $f_{01}$ with total pulse duration $t_{01}$ and the other π pulse at $f_{12}$ with duration $t_{12.}$ In the combined pulse, the time delay between the peaks of two pulses is $t_{\text{delay}}$. A pump signal at 13 dBm is kept on at Port 1 throughout the process. (**c**) Measured and simulated forward (backward) power at $f_2$ ($f_0$) as a function of delay time $t_{\text{delay}}$, when the resonator is pumped at $f_0$ ($f_2$). (**d**) Measured and simulated nonreciprocity as a function of delay time $t_{\text{delay}}$. $f_{01}$ = 909 kHz, $f_{12}$ = 1240 kHz, $t_{01}$ = 14 μs, $t_{12}$ = 30 μs, $T$ = 80 μs, other device parameters are listed in **Table. S1**.

In experiment, we use a continuous wave input and modulation sequences at a repetition rate of $1/T$ (**Fig. 4(b)**). In addition, Blackman-Harris windows are used to shape the modulation pulses to suppress the spectral leakage due to the sidebands of modulation signals. The effective lengths of two modulation pulses $\hat{t}_{01}$ and $\hat{t}_{12}$ are set to $\frac{\pi}{\Omega_1}$ and $\frac{\pi}{\Omega_2}$, respectively, i.e., $\pi$ pulses for each transition (see **Sec. 4** of **Supplemental Material**). We tune the delay $t_{\text{delay}}$ to control nonreciprocity, that is, the ratio of the forward (from Mode 0 to Mode 2) and backward (from Mode 2 to Mode 0) converted power (**Fig. 4(c)-4(d)**). We achieve a maximum nonreciprocity of 20.1 dB at $t_{\text{delay}}$=11 μs from $f_0$ to $f_2$ and 13.8 dB at $t_{\text{delay}}$= -9 μs from $f_2$ to $f_0$. On the other hand, the nonreciprocity vanishes at $t_{\text{delay}}$= 0 as the modulation signal becomes time reversal symmetric. Our numerical simulations that are based on the corresponding time-dependent Hamiltonian agree with the experimental results (see **Sec. 6** and **Sec. 7** of **Supplemental Material**).

In our configuration, the nonreciprocity of our system depends on (1) the temporal modulation sequence, (2) the loss of the modes, and (3) the difference between the spectral spacing $f_{12}$ -$f_{01}$. For a given modulation period $T$, there is an optimal $Q$ to reach maximum nonreciprocity (**Fig. S7(a)**). Modes with longer lifetime (higher $Q$) require a longer modulation period $T$ for maximum nonreciprocity (**Fig. S7(b)**). We note that although this modulation-pulse-induced non-reciprocal frequency conversion in a single device could not continuously convert inputs, we could employ multiple devices in parallel with a switch to achieve that, similar to non-reciprocal devices based on spatiotemporal modulation [51].

***Summary and discussion*** – We demonstrate an integrated platform for cavity electro-acoustic dynamics. The demonstrated non-reciprocal frequency conversion scheme provides a programmable method to achieve magnetic-free non-reciprocal acoustic devices. Our platform could find immediate applications in signal processing, acoustic computing [34], and topological acoustics [52,53]. We note that our present demonstrations of modulations are performed using coherent sources but with a potential to manipulate single acoustic phonons. The electro-acoustic modulation remains effective down to the single-phonon level at cryogenic temperatures [44]. In addition, better performance metrics, such as higher $Q$ factors, are expected when cooling the acoustic-wave device down to cryogenic temperatures [46,47]. Single phonons can be generated from optomechanical interactions [54], the coupling with superconducting qubits [9,55], and, potentially, mechanical spontaneous down conversion (SPDC).

***Acknowledgements*** – We thank Rohde & Schwarz for support with microwave instrumentation. Device fabrication was conducted at the Center for Nanophase Materials Sciences (CNMS2022-B-01473, CNMS2024-B-02643, L.S.), which is a US Department of Energy Office of Science User Facility. Research was partially supported by the Air Force Office of Scientific Research (AFOSR) under Grant Number W911NF-23-1-0235 (L.S.). Development of the optical vibrometer was partially supported by the Defense Advanced Research Projects Agency (DARPA) OPTIM program under contract HR00112320031 (L.S. and Y.Z.). Design and simulation of the acoustic device were partially supported by DARPA SynQuaNon DO program under Agreement HR00112490314 (L.S.). The views and conclusions contained in this document are those of the authors and do not necessarily reflect the position or the policy of the United States Government. No official endorsement should be inferred. Approved for public release.

***Author contributions*** – L.S. conceptualized the idea. J.J. and L.S. developed the theory. J.J. performed numerical simulations. J.J. and L.S. designed the chip. J.J. and Z.X. fabricated the devices; D.P.B. and I.I.K. developed fabrication recipes. J.J., L.S., and L.Z. conducted the device measurements, J.G.T. and Y.Z. performed optical vibrometer measurements. L.Z., L.S. and L.J built the custom high-voltage amplifier. J.J., A.G.P, and L.S. analyzed and interpreted the results. J.J. prepared the draft of manuscript. L.S. revised the manuscript with contributions from all authors. L.S. supervised the project.

***Competing interests*** – The authors declare no competing interests.

***Data availability*** – The data that supports the findings of this article cannot be made publicly available because they contain commercially sensitive information. The data are available upon reasonable request from the authors.

# Supplemental Materials for

## On-chip cavity electro-acoustics using lithium niobate phononic-crystal resonators

Jun Ji, Joseph G. Thomas, Zichen Xi, Liyang Jin, Dayrl P. Briggs, Ivan I. Kravchenko,
Arya G. Pour, Liyan Zhu, Yizheng Zhu, and Linbo Shao*

Corresponding author: shaolb@vt.edu

**The PDF file includes:**

Supplementary Section S1-S8
Figs. S1 to S8
Tables S1
Caption for Movie S1

**Other Supplementary Materials for this manuscript include the following:**

Movies S1

The supplementary material is organized as follows. **Section S1** describes the design principles and fabrication process of the phononic crystal (PnC) resonator. **Section S2** presents the experimental setups used for the characterization of the PnC resonators and for demonstrating both the spectral and temporal dynamics in our cavity electro-acoustic platform. **Section S3** describes the design of the high-voltage amplifier used to implement the electrical modulation. **Section S4** discusses the effective pulse length in the temporal dynamics measurements. **Sections S5 and S6** present the analysis of the two-level and three-level systems using equations of motion and time-dependent Hamiltonians. **Section S7** provides numerical simulations of the nonreciprocal frequency conversion in the three-level system. **Section S8** describes the mechanism of the electro-acoustic interaction.

## S1. Device design and fabrication

We design the device is on an *X*-cut LN substrate with a SiN thin film on top. We pattern the SiN layer to define the PnC resonator and other acoustic-wave structures. Monolithic interdigital transducers (IDTs) are used to excite and detect SAW. The pitch of IDT electrodes is 1.651 μm, which corresponds to the half wavelength of acoustic wave at 1.005 GHz. The width of the IDT is 50 μm. The thickness of aluminum is 100 nm. The number of IDT pairs is 35. A 300-μm-long taper is used to connect the IDT region with the 6-μm-wide acoustic waveguide.

The PnC resonator is orientated in a 30° angle with respect to the crystal *Z* axis to minimize the loss. The resonator is composed of taper, mirror, and cavity parts (**Fig. S1(a)**). The pitches of SiN pillars in the cavity and mirror region are 1.71 and 1.68 μm, respectively. The distribution of pitch, length, and width of pillars in the resonator are shown in **Fig. S1(b)**. The number of pillars in the waveguide-to-mirror tapering region, the mirror region, the mirror-to-cavity tapering region, and the cavity region are 10, 40, 30, and 350. Upper band modes of the cavity PnC fall in the middle of the bandgap of the mirror PnC (**Fig. S1(c)**), enabling the formation of localized cavity modes. The strong dispersion of the upper band at *X* point ($k$ =1) leads to the non-uniform mode spacings among acoustic modes. Simulated SAW modes of PnC with π phase of the unit cells are in **Fig. S1(d)**. Bandgaps of the cavity and mirror PnC are observed from the measured transmission spectra in **Fig. S1(e),** with frequency ranges that agree well with the simulated ones. The measured resonant frequencies of our PnC resonator are located close to the upper band edge of the cavity PnC. We further visualize out-of-plane displacement profiles of the three modes in **Fig. S1(f) - S1 (h)**, using our in-house microwave-frequency optical vibrometer. Within the cavity region, the displacement profile of Mode $\boldsymbol{n}$ (with $\boldsymbol{n}$ = 0,1,2) exhibits $\boldsymbol{n}$ nodes.

The fabrication processes of our devices are summarized as following. A 300-nm-thick SiN layer is deposited using plasma-enhanced chemical vapor deposition (PECVD) on an *X*-cut LN substrate. The SiN layer is patterned by electron-beam lithography (EBL) using polymethyl methacrylate (PMMA) resist and etched by reactive-ion etching using carbon tetrafluoride, sulfur hexafluoride, and trifluoromethane gases. The metal layer is patterned by EBL using PMMA resist. A 100-nm-thick aluminum layer is deposited by electron-beam evaporation, followed by lift-off in N-Methyl-2-pyrrolidone (NMP).

## S2. Device characterizations and experimental setups

In spectral domain measurements (**Figs. 1** and **2** in the main text and **Figs. S3-S5**), the *S*-parameter spectra (**Fig. S8(a)**) are measured using a vector network analyzer (Keysight P5000A). The continuous-wave modulation signal is generated by a function generator (Rigol DG2102) and then amplified by a custom high-voltage amplifier.

In the Rabi oscillation measurement (**Fig. 3** in the main text), we need a microwave pulse to pump the resonator, a modulation pulse, and capture of real-time output waveforms (**Fig. S8(b)**). Our microwave signal generator (RIGOL DSG836A) operates in pulse mode to pump the resonator. The electrical modulation pulse is produced by a function generator (the function generator of Rohde & Schwarz MXO4) and then amplified by a custom high-voltage amplifier. A trigger-out signal from the microwave signal generator synchronizes the function generator, ensuring a precise alignment between the pump and modulation pulses. The output GHz signal is first down-converted by an in-phase and quadrature ($I/Q$) demodulator (Analog Devices ADL5380) to a MHz frequency signal. Another microwave generator (RIGOL DSG821) serves as a local oscillator in the down conversion, with its frequency set to 1 MHz below $f_0$. The 1 MHz offset avoids electromagnetic interference between the down-converted signal and the high-voltage electrical modulation signal. The down-converted MHz signal is captured by an oscilloscope (Rohde & Schwarz RTO6). To extract the real-time amplitude of Mode 0 and Mode 1 (**Fig. 3(c)** in the main text), the acquired time-domain data is first digitally rotated back into the frame of their resonant frequency $f_0$ and $f_1$, respectively, and then passed through a 200 kHz low-pass filter.

In the non-reciprocal frequency conversion measurement (**Fig. 4** in the main text), the spectra are read (**Fig. S8(c)**) by the spectrum analyzer module in vector network analyzer (Keysight P5004A). The forward power is the output power measured at $f_2$ when the resonator is pumped at $f_0$. The backward power is the output power measured at $f_0$ when the resonator is pumped at $f_2$. The nonreciprocity is the ratio between the forward power and the backward power.

## S3. The custom high-voltage amplifier

In our implementation, two operational amplifiers (APEX PA96) are configured in complementary inverting and non-inverting modes, yielding output signals that are 180° out of phase. This arrangement establishes a bridge-tied load (BTL) drive, effectively doubling the differential output voltage compared to a single-ended configuration and thereby enhancing the available drive capability. A regulated power supply section is designed to provide low-voltage rails (up to ±15 V) and a high DC supply voltage (300 V) for the op-amps. Limited by the slew rate of operational amplifiers, the 300 Vpp output voltage at 1 MHz is a triangle-like wave. High-order harmonics in the triangle-like wave are relatively weak and barely induce an electro-acoustic effect on the devices since they do not align with the differences in modal frequencies.

## S4. Effective length of the modulation pulse

The effective pulse length is defined as $\hat{t} = \frac{\int_0^T W_{\mathrm{bh}}(t)\mathrm{d}t}{\max(W_{\mathrm{bh}}(t))}$, where $W_{\mathrm{bh}}$ is Blackman-Harris window. The effective pulse length in **Fig. 4** in the main text is 5 μs and 10 μs for pulse 1 and pulse 2, respectively. They correspond to π pulse modulations at Rabi frequencies of 96 kHz and 47 kHz (**Table. S1**).

## S5. Dynamics of electro-acoustic cavity with two acoustic modes

The equations of motion of our electro-acoustic cavity with two acoustic modes $a_0$ and $a_1$ under a continuous wave excitation at $\omega_{in}$ are described in the lab frame as:

$$i\frac{d\vec{a}}{dt} = H_{\mathrm{lab}}\vec{a} - i\overrightarrow{\alpha_{\mathrm{in}}}, \tag{S1}$$

where the state vector $\vec{a} = (a_0, a_1)^T$ and the input vector $\overrightarrow{\alpha_{\mathrm{in}}} = \alpha_{\mathrm{in}}e^{-i\omega_{\mathrm{in}}t}(\sqrt{\gamma_{e0}}, \sqrt{\gamma_{e1}})^T$. The Hamiltonian $H_{lab}$ is described by,

$$H_{\mathrm{lab}} = H_0 + H_{\mathrm{int}}(t), \tag{S2}$$

$$H_0 = \begin{pmatrix} \omega_0 - i\frac{\gamma_0}{2} & 0 \\ 0 & \omega_1 - i\frac{\gamma_1}{2} \end{pmatrix}, \tag{S3}$$

$$H_{\mathrm{int}}(t) = \begin{pmatrix} 0 & \mu_1 V_{\mathrm{mod}}\cos(\omega_{\mathrm{mod}}t) \\ \mu_1 V_{\mathrm{mod}}\cos(\omega_{\mathrm{mod}}t) & 0 \end{pmatrix}, \tag{S4}$$

where $\omega_n$, $\gamma_n$, and $\gamma_{en}$ are the resonant frequency, total loss rate, and external coupling rate of Mode $n$, respectively. $n = 0$ and 1. $H_{mod}$ is the coupling induced by the electrical modulation at the frequency $\omega_m$ with a coupling strength of $\Omega_1 = \mu_1 V_{mod}$ between Mode 0 and Mode 1. $V_{mod}$ is the amplitude of the electrical modulation. For simplicity, we apply a unitary transformation $U$,

$$U = \begin{pmatrix} e^{i\omega_{\mathrm{in}}t} & 0 \\ 0 & e^{i(\omega_{\mathrm{in}}+\omega_{\mathrm{mod}})t} \end{pmatrix}. \tag{S5}$$

**Equation S1** becomes,

$$i\frac{d\vec{c}}{dt} = H_{\mathrm{rot}}\vec{c} - i\overrightarrow{c_{\mathrm{in}}}, \tag{S6}$$

where the rotated state vector $\vec{c} = U\vec{a} = (c_0, c_1)^T$ and the rotated input vector $\overrightarrow{c_{\mathrm{in}}} = U\overrightarrow{\alpha_{\mathrm{in}}}$. The rotated Hamiltonian $H_{rot}$ with the rotation wave approximation is described by,

$$H_{\mathrm{rot}}(t) = UH_{\mathrm{lab}}U^\dagger - iU\frac{dU^\dagger}{dt} = \begin{pmatrix} \omega_0 - \omega_{\mathrm{in}} - i\frac{\gamma_0}{2} & \frac{\mu_1 V_m}{2} \\ \frac{\mu_1 V_m}{2} & \omega_1 - \omega_{\mathrm{in}} - \omega_{\mathrm{m}} - i\frac{\gamma_1}{2} \end{pmatrix}. \tag{S7}$$

**Equation S6** becomes,

$$\dot{c}_0 = (i\Delta - \gamma_0/2)c_0 - i\frac{\Omega_1}{2}c_1 - \sqrt{\gamma_{e0}}c_{\mathrm{in}}, \tag{S8}$$

$$\dot{c}_1 = (i\Delta + i\delta - \gamma_1/2)c_1 - i\frac{\Omega_1}{2}c_0 - \sqrt{\gamma_{e1}}c_{\mathrm{in}}e^{i\omega_{\mathrm{m}}t}, \tag{S9}$$

where $\Delta = \omega_{in} - \omega_0$ is the excitation detuning with respect to Mode 0 and $\delta = \omega_m - \omega_{01}$ is the detuning of electrical modulation frequency with respect to the difference of mode frequencies ($\omega_{01} = \omega_1 - \omega_0$).

**Equations S8** and **S9** are nonlinear and coupled equations, we can solve them by expanding $c_0$ and $c_1$ into three frequency components – the excited frequency and two first-order sideband frequencies. That is,

$$c_0 = c_{0,0} + c_{0,-\omega_\mathrm{m}} e^{-i\omega_\mathrm{m} t} + c_{0,\omega_\mathrm{m}} e^{i\omega_\mathrm{m} t} + \cdots, \tag{S10}$$

$$c_1 = c_{1,0} + c_{1,-\omega_\mathrm{m}} e^{-i\omega_\mathrm{m} t} + c_{1,\omega_\mathrm{m}} e^{i\omega_\mathrm{m} t} + \cdots. \tag{S11}$$

Plugging **Eqs. S10** and **S11** into **Eqs. S8** and **S9**, we get

$$c_{0,0} = \frac{\sqrt{\gamma_{e0}}\alpha_\mathrm{in}}{(i\Delta - \gamma_0/2) + \dfrac{\Omega_1^2/4}{i(\Delta + \delta) - \gamma_1/2}}, \tag{S12}$$

$$c_{1,0} = \frac{i\sqrt{\gamma_{e0}}\alpha_\mathrm{in}\,\Omega_1/2}{(i\Delta - \gamma_0/2)[i(\Delta + \delta) - \gamma_1/2] + \Omega_1^2/4}, \tag{S13}$$

$$c_{0,\omega_\mathrm{m}} = \frac{i\sqrt{\gamma_{e1}}\alpha_\mathrm{in}\,\Omega_1/2}{[i(\Delta + \delta - \omega_\mathrm{m}) - \gamma_1/2][i(\Delta - \omega_\mathrm{m}) - \gamma_0/2] + \Omega_1^2/4}, \tag{S14}$$

$$c_{1,\omega_\mathrm{m}} = \frac{\sqrt{\gamma_{e1}}\alpha_\mathrm{in}}{[i(\Delta + \delta - \omega_\mathrm{m}) - \gamma_1/2] + \dfrac{\Omega_1^2/4}{i(\Delta - \omega_\mathrm{m}) - \gamma_0/2}}, \tag{S15}$$

$$c_{0,-\omega_\mathrm{m}} = c_{1,-\omega_\mathrm{m}} = 0. \tag{S16}$$

The output at the lab frame is:

$$a_\mathrm{out} = \left(\sqrt{\gamma_{e0}}c_0 + \sqrt{\gamma_{e1}}c_1 e^{-i\omega_\mathrm{m} t}\right)e^{-i\omega_\mathrm{in} t}. \tag{S17}$$

The transmission at the excitation frequency is:

$$a_\mathrm{out,0} = \left(\sqrt{\gamma_{e0}}c_{0,0} + \sqrt{\gamma_{e1}}c_{1,\omega_\mathrm{m}}\right)e^{-i\omega_\mathrm{in} t}. \tag{S18}$$

The amplitude of the transmission spectrum $S_{21}$ can be calculated by:

$$|S_{21}|^2 = \left|\frac{a_\mathrm{out,0}}{\alpha_\mathrm{in}}\right|^2 = \left|\frac{\gamma_{e0}}{(i\Delta - \gamma_0/2) + \dfrac{\Omega_1^2/4}{i(\Delta + \delta) - \gamma_1/2}} + \frac{\gamma_{e1}}{[i(\Delta + \delta - \omega_\mathrm{m}) - \gamma_1/2] + \dfrac{\Omega_1^2/4}{i(\Delta - \omega_\mathrm{m}) - \gamma_0/2}}\right|^2. \tag{S19}$$

The device parameters $f_0$, $f_1$, $\gamma_0$, $\gamma_1$, $\gamma_{0\mathrm{ext}}$, and $\gamma_{1\mathrm{ext}}$ are extracted from $S_{21}$ measurements without the electrical modulation in **Fig. 1(c)** in the main text. Then, the coupling strength $\Omega_1$ is extracted by matching measured transmission spectra in **Fig. S4** to the analytical solution (**Eq. S19**). A similar procedure applies to the estimation of the coupling strength $\Omega_2$ in **Fig. S5.**

## S6. Dynamics of electro-acoustic cavity with three spectrally unevenly-spaced acoustic modes

The equations of motion of our electro-acoustic cavity with three acoustic modes $a_0$, $a_1$, and $a_2$ under a continuous wave excitation at $\omega_{in}$ are described in the lab frame as:

$$i\frac{d\vec{a}}{dt} = H_{\mathrm{lab}}\vec{a} - i\overrightarrow{\alpha_{\mathrm{in}}}, \tag{S20}$$

where the state vector $\vec{a} = (a_0, a_1, a_2)^T$ and the input vector $\overrightarrow{\alpha_{\mathrm{in}}} = \alpha_{\mathrm{in}}e^{-i\omega_{\mathrm{in}}t}(\sqrt{\gamma_{e0}}, \sqrt{\gamma_{e1}}, \sqrt{\gamma_{e2}})^T$. The periodic Hamiltonian $H_{lab}$ within a period of $T$ is described by,

$$H_{\mathrm{lab}} = H_{\mathrm{idle,lab}} + H_{\mathrm{p,lab}}(t) + H_{\mathrm{q,lab}}(t), \tag{S21}$$

$$H_{\mathrm{idle,lab}} = \begin{pmatrix} \omega_0 - i\frac{\gamma_0}{2} & 0 & 0 \\ 0 & \omega_1 - i\frac{\gamma_1}{2} & 0 \\ 0 & 0 & \omega_2 - i\frac{\gamma_2}{2} \end{pmatrix}, \tag{S22}$$

$$H_{\mathrm{p,lab}}(t) = \chi_{(\tau_{\mathrm{pump}},\tau_{\mathrm{pump}}+\tau_{\mathrm{p}})}(t)\begin{pmatrix} 0 & \frac{\mu_1}{2}V_{\mathrm{p}}e^{i\omega_{\mathrm{m1}}t} & 0 \\ \frac{\mu_1}{2}V_{\mathrm{p}}e^{-i\omega_{\mathrm{m1}}t} & 0 & \frac{\mu_2}{2}V_{\mathrm{p}}e^{i\omega_{\mathrm{m1}}t} \\ 0 & \frac{\mu_2}{2}V_{\mathrm{p}}(t)e^{-i\omega_{\mathrm{m1}}t} & 0 \end{pmatrix}, \tag{S23}$$

$$H_{\mathrm{q,lab}}(t) = \chi_{(\tau_{\mathrm{pump}}+\tau_{\mathrm{p}},\tau_{\mathrm{pump}}+\tau_{\mathrm{p}}+\tau_{\mathrm{q}})}(t)\begin{pmatrix} 0 & \frac{\mu_1}{2}V_{\mathrm{q}}e^{i\omega_{\mathrm{m2}}t} & 0 \\ \frac{\mu_1}{2}V_{\mathrm{q}}e^{-i\omega_{\mathrm{m2}}t} & 0 & \frac{\mu_2}{2}V_{\mathrm{q}}e^{i\omega_{\mathrm{m2}}t} \\ 0 & \frac{\mu_2}{2}V_{\mathrm{q}}e^{-i\omega_{\mathrm{m2}}t} & 0 \end{pmatrix}, \tag{S24}$$

$$\chi_{(a,b)}(t) = \begin{cases} 1, & a \le t < b, \\ 0, & 0 \le t < a \text{ or } b \le t < T. \end{cases} \tag{S25}$$

Here, $H_{p,lab}$ ($H_{q,lab}$) describes the coupling due to an electrical modulation pulse $p$ ($q$) at the frequency of $\omega_{m1} = \omega_1 - \omega_0$ ($\omega_{m2} = \omega_2 - \omega_1$) with a coupling strength of $\Omega_1 = \mu_1 V$ between Mode 0 and Mode 1 ($\Omega_2 = \mu_2 V$ between Mode 1 and Mode 2). Rotation wave approximation is applied. $\tau_{pump}$ is the time to pump the mode, $\tau_p$ ($\tau_q$) is the time for the electrical modulation pulse $p$ ($q$). Here, modulation pulses $p$ and $q$ are considered as completely non-overlapping for physical insights. The overlapping is considered in numerical simulations.

For simplicity, we apply a unitary transformation $U$,

$$U = \begin{pmatrix} e^{i\omega_0 t} & 0 & 0 \\ 0 & e^{i\omega_1 t} & 0 \\ 0 & 0 & e^{i\omega_2 t} \end{pmatrix}. \tag{S26}$$

**Equation S21** becomes,

$$i\frac{d\vec{c}}{dt} = H_{\mathrm{rot}}\vec{c} - i\overrightarrow{c_{\mathrm{in}}}, \tag{S27}$$

where the rotated state vector $\vec{c} = U\vec{\alpha} = (c_0, c_1, c_2)^T$ and the rotated input vector $\overrightarrow{c_{\rm in}} = U\overrightarrow{\alpha_{\rm in}}$. The Hamiltonian $H_{\rm rot}$ is described by,

$$H_{\rm rot}(t) = UH_{\rm lab}U^{\dagger} - iU\frac{dU^{\dagger}}{dt} = H_{\rm idle} + H_{\rm p}(t) + H_{\rm q}(t), \tag{S28}$$

$$H_{\rm idle} = \begin{pmatrix} -i\frac{\gamma_0}{2} & 0 & 0 \\ 0 & -i\frac{\gamma_1}{2} & 0 \\ 0 & 0 & -i\frac{\gamma_2}{2} \end{pmatrix}, \tag{S29}$$

$$H_{\rm p}(t) = \chi_{(\tau_{\rm pump}, \tau_{\rm pump}+\tau_{\rm p})}(t)\begin{pmatrix} 0 & \mu_1\cos(\omega_{\rm m1}t)\,V_{\rm p}e^{-i\omega_{01}t} & 0 \\ \mu_1\cos(\omega_{\rm m1}t)\,V_{\rm p}e^{i\omega_{01}t} & 0 & \mu_2\cos(\omega_{\rm m1}t)\,V_{\rm p}e^{-i\omega_{12}t} \\ 0 & \mu_2\cos(\omega_{\rm m1}t)\,V_{\rm p}e^{-i\omega_{12}t} & 0 \end{pmatrix}, \tag{S30}$$

$$H_{\rm q}(t) = \chi_{(\tau_{\rm pump}+\tau_{\rm p}, \tau_{\rm pump}+\tau_{\rm p}+\tau_{\rm q})}(t)\begin{pmatrix} 0 & \mu_1\cos(\omega_{\rm m2}t)\,V_{\rm q}e^{-i\omega_{01}t} & 0 \\ \mu_1\cos(\omega_{\rm m2}t)\,V_{\rm q}e^{-i\omega_{01}t} & 0 & \mu_2\cos(\omega_{\rm m2}t)\,V_{\rm q}e^{-i\omega_{12}t} \\ 0 & \mu_2\cos(\omega_{\rm m2}t)\,V_{\rm q}e^{i\omega_{12}t} & 0 \end{pmatrix}. \tag{S31}$$

Over one period, the propagator is,

$$U(T) = U_{\rm idle}U'_{\rm q}(\tau_{\rm q})U'_{\rm p}(\tau_{\rm p}) = U_{\rm idle}e^{-i\int_{\tau_{\rm pump}+\tau_{\rm p}}^{\tau_{\rm pump}+\tau_{\rm p}+\tau_{\rm q}}H_{\rm q}(t){\rm d}t}e^{-i\int_{\tau_{\rm pump}}^{\tau_{\rm pump}+\tau_{\rm p}}H_{\rm p}(t){\rm d}t}. \tag{S32}$$

Note that $H_{\rm p}$ and $H_{\rm q}$ do not commute, thus

$$U(T) \neq U_{\rm idle}e^{-i\int_{\tau_{\rm pump}}^{\tau_{\rm pump}+\tau_{\rm p}+\tau_{\rm q}}\left(H_{\rm p}(t)+H_{\rm q}(t)\right){\rm d}t}. \tag{S33}$$

In general cases that two temporal electrical modulations have a relative delay $t_{\rm delay} \neq 0$, the broken time reversal symmetry induces nonreciprocity (**Fig. 4(c)** in the main text).

In a parameter region that $f_{12} \gg f_{01}$, detuned transitions are inefficient and neglected compared with on-resonance transitions. **Equations S30** and **S31** become,

$$H_{\rm p}(t) = \chi_{(\tau_{\rm pump}, \tau_{\rm pump}+\tau_{\rm p})}(t)\begin{pmatrix} 0 & \frac{\mu_1 V_p}{2} & 0 \\ \frac{\mu_1 V_p}{2} & 0 & 0 \\ 0 & 0 & 0 \end{pmatrix}, \tag{S34}$$

$$H_{\rm q}(t) = \chi_{(\tau_{\rm pump}+\tau_{\rm p}, \tau_{\rm pump}+\tau_{\rm p}+\tau_{\rm q})}(t)\begin{pmatrix} 0 & 0 & 0 \\ 0 & 0 & \frac{\mu_2 V_q}{2} \\ 0 & \frac{\mu_2 V_q}{2} & 0 \end{pmatrix}. \tag{S35}$$

For the transition from Mode 0 to Mode 1, the 2×2 propagator is,

$$U'_{\rm p}(\tau_{\rm p}) = e^{-\frac{\bar{\gamma}_{01}\tau_{\rm p}}{2}}\left[\cos\frac{\Omega_{R1}\tau_{\rm p}}{2}I - i\frac{\sin\left(\Omega_{R1}\tau_{\rm p}/2\right)}{\Omega_{R1}}(i\delta\gamma_{01}\sigma_z + \Omega_1\sigma_x)\right], \tag{S36}$$

where $\bar{\gamma}_{01} = (\gamma_0 + \gamma_1)/2$, $\delta\gamma_{01} = (\gamma_1 - \gamma_0)/2$, and $\Omega_{R1} = \sqrt{\Omega_1^2 - \delta\gamma_{01}^2}$, $\sigma_x$ and $\sigma_z$ are Pauli matrices. Embedded in the 3×3 propagator, it becomes,

$$U_{\mathrm{p}} = \begin{pmatrix} [U'_{\mathrm{p}}]_{00} & [U'_{\mathrm{p}}]_{01} & 0 \\ [U'_{\mathrm{p}}]_{10} & [U'_{\mathrm{p}}]_{11} & 0 \\ 0 & 0 & e^{-\frac{\gamma_2 \tau_{\mathrm{p}}}{2}} \end{pmatrix}. \tag{S37}$$

Similarly, the 2×2 propagator for the transition from Mode 1 to Mode 2 is,

$$U'_q(\tau_{\mathrm{q}}) = e^{-\frac{\bar{\gamma}_{12}\tau_{\mathrm{q}}}{2}} \left[ \cos\frac{\Omega_{R2}\tau_{\mathrm{q}}}{2} I - i\frac{\sin(\Omega_{R2}\tau_{\mathrm{q}}/2)}{\Omega_{R2}} (i\delta\gamma_{12}\sigma_z + \Omega_2\sigma_x) \right], \tag{S38}$$

where $\bar{\gamma}_{12} = (\gamma_1 + \gamma_2)/2$, $\delta\gamma_{12} = (\gamma_2 - \gamma_1)/2$, and $\Omega_{R2} = \sqrt{\Omega_2^2 - \delta\gamma_{12}^2}$. Embedded in the 3×3 propagator, it becomes,

$$U_{\mathrm{q}} = \begin{pmatrix} e^{-\frac{\gamma_0 \tau_{\mathrm{q}}}{2}} & 0 & 0 \\ 0 & [U'_q]_{00} & [U'_q]_{01} \\ 0 & [U'_q]_{10} & [U'_q]_{11} \end{pmatrix}. \tag{S39}$$

The combined propagator is,

$$U_{\mathrm{c}} = U_{\mathrm{q}}U_{\mathrm{p}} = \begin{pmatrix} e^{-\frac{\gamma_0 \tau_{\mathrm{q}}}{2}}[U'_{\mathrm{p}}]_{00} & e^{-\frac{\gamma_0 \tau_{\mathrm{q}}}{2}}[U'_{\mathrm{p}}]_{01} & 0 \\ [U'_{\mathrm{p}}]_{10}[U'_q]_{00} & [U'_{\mathrm{p}}]_{11}[U'_q]_{00} & e^{-\frac{\gamma_2 \tau_{\mathrm{p}}}{2}}[U'_q]_{01} \\ [U'_{\mathrm{p}}]_{10}[U'_q]_{10} & [U'_{\mathrm{p}}]_{11}[U'_q]_{10} & e^{-\frac{\gamma_2 \tau_{\mathrm{p}}}{2}}[U'_q]_{11} \end{pmatrix}. \tag{S40}$$

In this case, our system features an infinite nonreciprocity: the forward frequency conversion efficiency is $[U'_{\mathrm{p}}]_{10}[U'_{\mathrm{q}}]_{10}$, when the resonator is pumped at Mode 0; the backward efficiency is 0, when pumped at Mode 2.

## S7. Numerical simulations of non-reciprocal frequency conversions

We numerically simulate the non-reciprocal frequency conversion process in **Fig. 4** in the main text and **Fig. S7** based on **Eq. S27**. In the simulation, we consider an overlap between two modulation pulses and Blackman-Harris window applied to the pulses. The modulation period $T$ is set as an integer multiple of $1/(f_2-f_0)$ to align the phase of the modulation signal between cycles.

Furthermore, we numerically explore the non-reciprocal frequency conversion using different intrinsic $Q$ factors of modes and different modulation periods $T$. Here, we introduce two dimensionless parameters: α simultaneously scales intrinsic $Q$ factors of all three modes; β changes the modulation period. The intrinsic $Q$ factor in simulation $Q'_{i,\mathrm{int}} = \alpha Q_{i,\mathrm{int}}$, where $Q_{i,\mathrm{int}}$ is the experimental intrinsic $Q$ factors, $i$ = 0, 1, and 2 for three modes. The modulation period in simulation $T' = \beta T$, where $T$ = 80 μs is the experimental modulation period. In the other word, the simulation with α = β = 1 matches the experiment in **Fig. 4** in the main text.

While keeping modulation period constant (β =1), we sweep the intrinsic $Q$ parameter α (**Fig. S7(a)**). When the modes are lossy represented by a small α (e.g., α = 0.1), the energy in modes decays faster than being transit to another modes. Consequently, both forward and backward power are low, leading to a small nonreciprocity. When the modes are high-$Q$ represented by a α larger than 100 (which could be realized at cryogenic temperature), the frequency conversion is efficient in both directions: Mode 0 is efficiently converted to Mode 2 through $f_{01}$ pulse and $f_{12}$ pulse within the same modulation period, while Mode 2 is efficiently converted to Mode 1 through $f_{21}$ in the first modulation period, and from Mode 1 to Mode 0 through $f_{01}$ in the subsequent modulation period. This leads to a small nonreciprocity. In this case, nonreciprocity is mainly determined by the lifetime of Mode 1 in respect to the idle time in the modulation period. In the case of β =1, the optimal intrinsic $Q$ factor parameter α is between 1 and 10.

For those modes with higher $Q$ factors, a lager modulation period is needed for nonreciprocity. We choose α = 20 as an example and sweep the modulation period parameter β (**Fig. S7(b)**). Nonreciprocity reaches a maximum of 51.6 dB at β = 3. When β < 3, the nonreciprocity becomes smaller with a smaller β. This is because the frequency conversion in the forward direction from Mode 0 to Mode 2 remains the same, while the frequency conversion in the backward direction from Mode 2 to Mode 0 becomes more efficient for a smaller β. When β > 3, nonreciprocity reaches a plateau. In this case, nonreciprocity is mainly determined by unwanted detuned transitions between modes. For α = 20 and β = 2, we could achieve a nonreciprocity of 40.4 and a forward power of -20.9 dBm. In this case, the cable-to-cable insertion loss is 33.9 dB, and the on-chip insertion loss is 22.9 dB considering 11 dB loss from the IDT pair.

## S8. The electro-acoustic interaction between acoustic modes

The nonlinear piezoelectric (electro-acoustic) effect describes how an applied electric field modifies the elasticity of material, and thus parametrically modulates the acoustic modes. When an electric field $E_k$ is applied, a change of elasticity matrix $\Delta c_{ij}$ is realized through the third-order piezoelectric tensor $D$ ($d_{kij}$), that is, $\Delta c_{ij} = d_{ijk} E_k$, where $i, j = 1,\ldots 6$, $k = 1$, 2 and 3.

The strength of electro-acoustic nonlinear interaction between two acoustic modes, which is represented by $\mu V$ in the time-dependent Hamiltonian **Eq. S4**, is determined by the integration of the two mechanical mode profiles and the dynamic modulation (the electric field distribution $E_k$ produced by the modulation electrode multiplied by the corresponding nonlinear piezoelectric coefficient),

$$\mu_{m,n} V = \frac{\sqrt{\omega_m \, \omega_n} \, \sum_{i,j} \int \, \Delta c_{ij} \epsilon_{i,m}^{*} \epsilon_{j,n} \, dr}{\sqrt{\sum_i \int \, c_{ii} \epsilon_{i,m}^{*} \epsilon_{i,m} dr \, \sum_i \int \, c_{ii} \epsilon_{i,n}^{*} \epsilon_{i,n} \, dr}}. \quad (S41)$$

Here, $\Delta c_{ij} = d_{ijk} E_k$, $\omega_m$ and $\omega_n$ is the eigenfrequency of mode $\boldsymbol{m}$ and $\boldsymbol{n}$, $\epsilon_{i,m}$and $\epsilon_{j,n}$denote the strain component $i$ and $j$ of acoustic modes $\boldsymbol{m}$ and $\boldsymbol{n}$. Matrix elements are summed following the Einstein summation convention. The applied modulation voltage satisfies $V = \int E_k \mathrm{d}l$ , corresponding to the line integration of the electrical field between the pair of modulation electrodes.

From this expression, the nonlinear interaction strengths are determined by (i) the electric field amplitude, either through a larger applied voltage or a smaller electrode gap, and (ii) optimizing the device orientation so that the dominant strain components of the acoustic modes couple to large nonlinear piezoelectric coefficients $d_{ijk}$associated with the applied electric field.

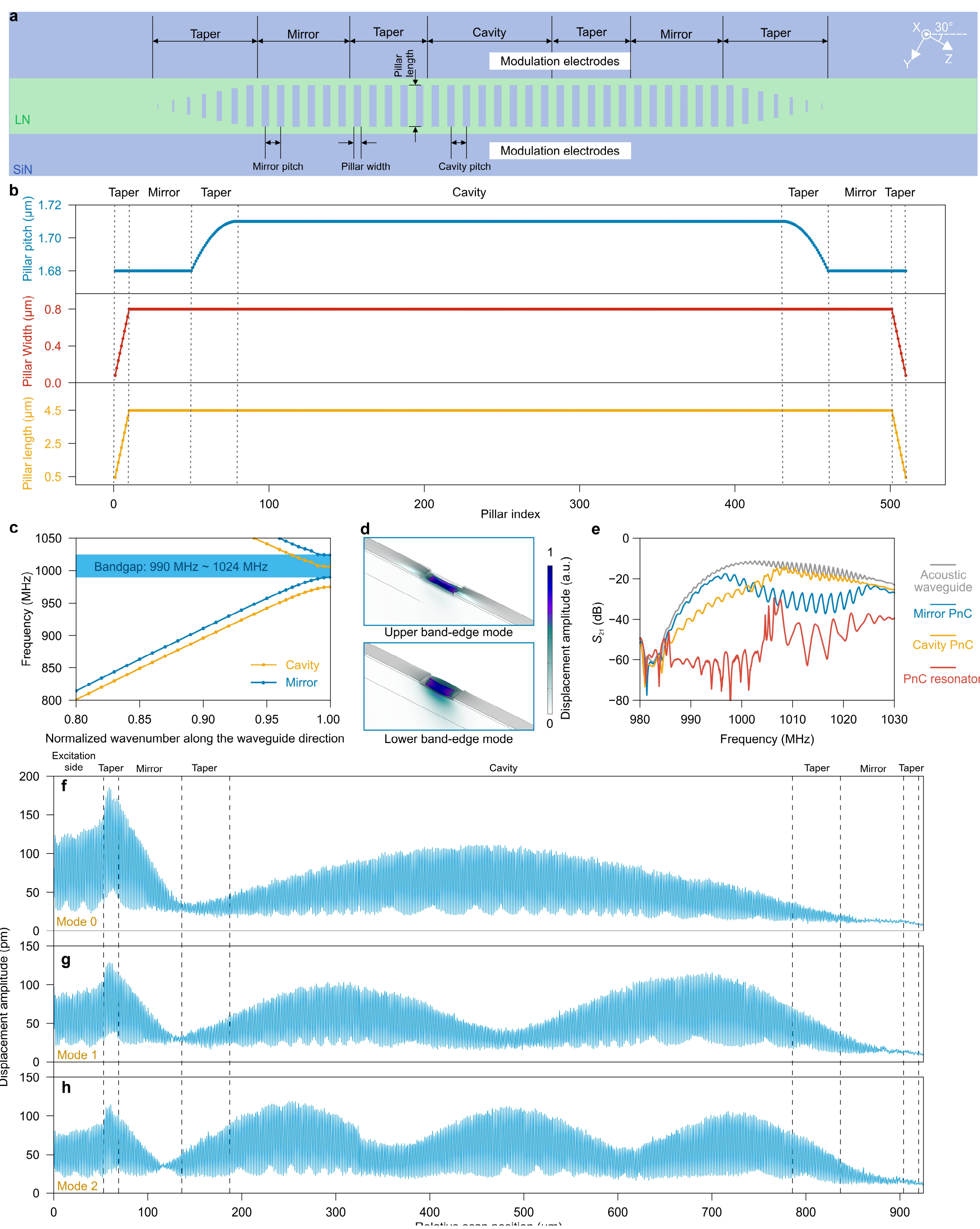

**Fig. S1. Design of our PnC resonator.** Full caption is on the next page.

**Fig. S1. Design of our PnC resonator.**
(**a**) The distribution of PnC pillars of our resonator. The PnC resonator is defined by a series of etched SiN pillars with different sizes in different segments. (**b**) The distribution of pitch, width, and length of SiN pillars. The pitch of the etched pillars in the cavity region is 1.71 μm and quadratically tapers to 1.68 μm in the mirror regions. The width (length) of the first 10 SiN pillar near IDTs linearly taper from 0.08 μm (0.5 μm) to 0.8 μm (4.5 μm) in the mirror region. Counting along the waveguide direction, the number of SiN pillars in different regions is: 10 (taper), 40 (mirror), 30 (taper), 350 (cavity), 30 (taper), 40 (mirror), 10 (taper). (**c**) Simulated phononic band structure of the cavity PnC and the mirror PnC. The mirror PnC forms a band gap ranging from 990 to 1,024 MHz; The cavity PnC forms a band gap ranging from 975 to 1,006 MHz. The upper band edge of the cavity PnC is aligned to the center of the band gap of the mirror PnC. (**d**) SAW modes profiles of the upper and lower band-edge modes with π phase of the unit cells. (**e**) Measured transmission $S_{21}$ spectra of the bare acoustic waveguide, 80-period mirror PhC, 80-period cavity PnC, and PnC resonator. Three high-$Q$ phononic modes are inside the bandgap of the mirror PnC and close to the upper band edge of the cavity PnC. The close-up view of the three modes is in **Fig. 1(c)** of main text. (**f**) to (**h**) Measured out-of-plane displacement profiles of the three modes, along the acoustic waveguide direction, using our in-house optical vibrometer. Within the cavity region, $\boldsymbol{n}$ nodes are observed in the displacement profile of mode $\boldsymbol{n}$, $\boldsymbol{n}$ = 0,1,2. Each mode is excited by applying a continuous-wave signal at the corresponding modal frequency at 10 dBm to the IDT at the left side.

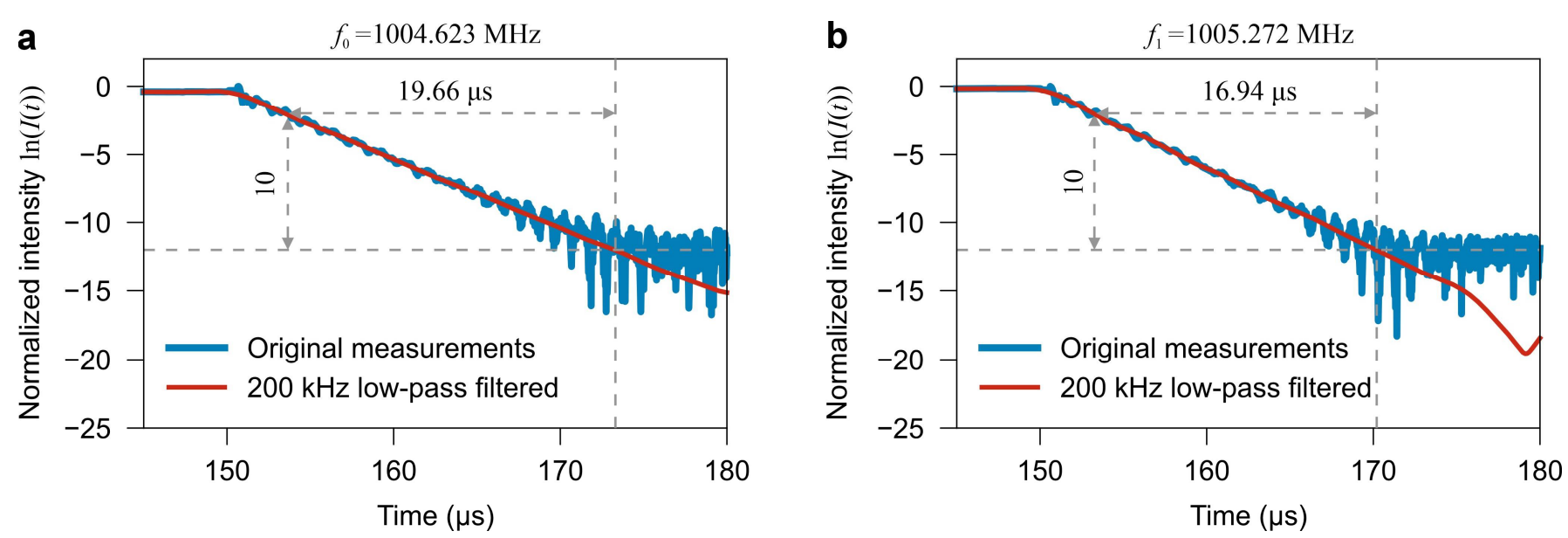

**Fig. S2. Ring-down measurements of our PnC resonator.** Measured normalized intensity curve $\ln(I(t))$ for (**a**) Mode 0 at 1,004.623 MHz and (**b**) Mode 1 at 1,005.272 MHz. $I(t) = |V(t)|^2/|V_{\max}|^2$, $V$ is the measured output voltage. The lifetime $\tau$, which is defined as the time taken for the energy to decay to 1/e, corresponds to the time for $\ln(I(t))$ to drop by 1. The measured lifetime $\tau$ is 1.966 μs and 1.694 μs for Mode 0 and 1, respectively. The measured $Q = 2\pi f\tau$ is 12,403 and 10,694 for Mode 0 and 1, respectively. The ripple in the original measurements is due to the slight excitation of neighboring modes when the pump signal is turned off sharply.

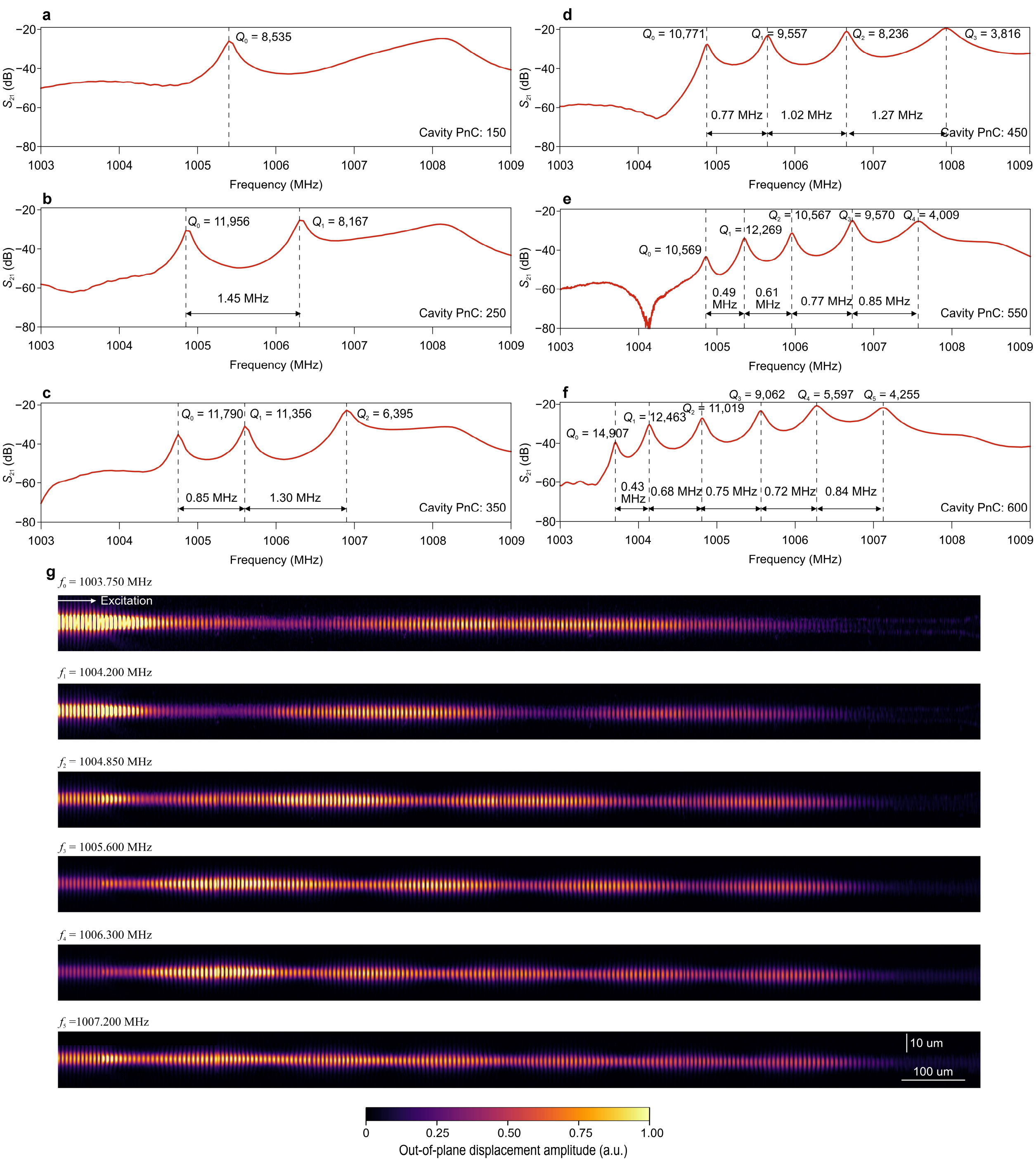

**Fig. S3. PnC resonators with engineered mode spacings.** Full caption is on the next page.

**Fig. S3. PnC resonators with engineered mode spacings.** As the length of the PnC cavity (equivalently, the number of PnC periods in the defect region) increases, both the number of high-$Q$ phononic modes increases and mode spacings decrease. For a fixed mirror design with 40 PnC periods on each side, the measured transmission spectra $S_{21}$ of PnC resonators show: (**a**) a single high-$Q$ mode for 150 cavity periods; (**b**) two high-$Q$ modes with a mode spacing of 1.45 MHz for 250 cavity periods; (**c**) three high-$Q$ modes with non-uniform spacings of 0.85 MHz and 1.30 MHz for 350 cavity periods; (**d**) four high-$Q$ modes with non-uniform spacings of 0.77 MHz, 1.02 MHz and 1.27 MHz for 450 cavity periods; (**e**) five high-$Q$ modes with non-uniform spacings of 0.49 MHz, 0.61 MHz, 0.77 MHz, and 0.85 MHz for 550 cavity periods; (**f**) six high-$Q$ modes with non-uniform spacings of 0.43 MHz, 0.68 MHz, 0.75 MHz, 0.72 MHz, and 0.84 MHz for 600 cavity periods. The same nominal design can yield slightly different spacings (e.g., panel (**c**) versus **Fig. 1(c)**) due to fabrication tolerances. (**g**) Measured 2D out-of-plane displacement profiles of these modes in (**f**), along the acoustic waveguide direction, using our in-house optical vibrometer. Within the cavity region, $\boldsymbol{n}$ nodes are observed in the displacement profile of mode $\boldsymbol{n}$, $\boldsymbol{n} = 0, 1, \ldots, 5$. Each mode is excited by applying a continuous-wave drive at the corresponding resonance frequency (10 dBm) to the input IDT on the left.

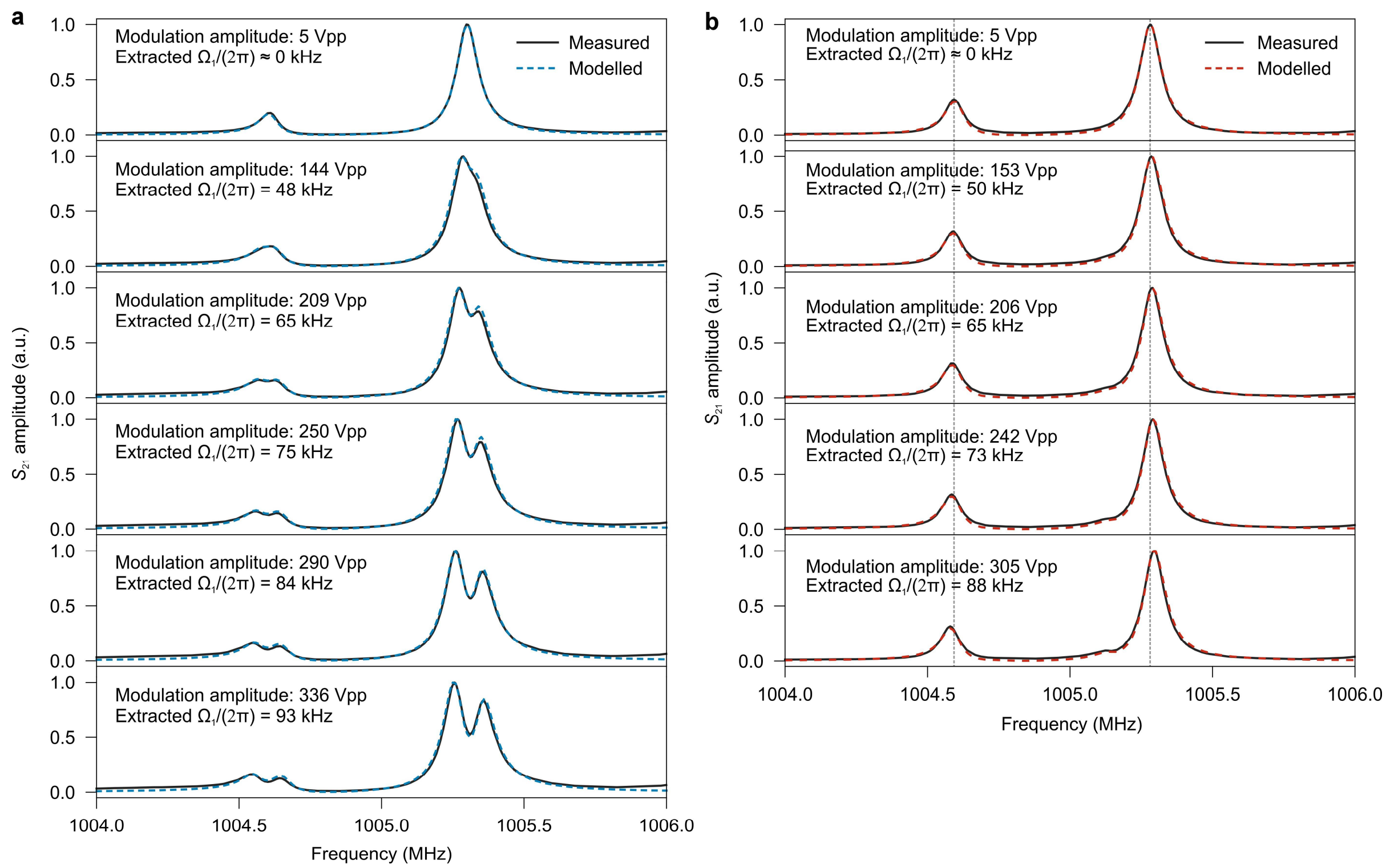

**Fig. S4. Calculated and measured transmission spectrum of electrically modulated acoustic modes.** (**a**) Calculated transmission spectrum using **Eq. S19** for the measured acoustic ATS in **Fig. 2a** in the main text. (**b**) Calculated transmission spectrum using **Eq. S19** for the measured acoustic a.c. Stark shifts in **Fig. 2b** in the main text. Extracted parameters: $f_0$ =1,004.600 MHz, $f_1$ = 1,005.280 MHz, $Q_0$ = 12,020, $Q_1$ = 10,600, $\gamma_0 = 2\pi f_0/Q_0$, $\gamma_1 = 2\pi f_1/Q_1$, $\gamma_{0ext} = \gamma_0/49$, $\gamma_{1ext} = \gamma_1/26.5$.

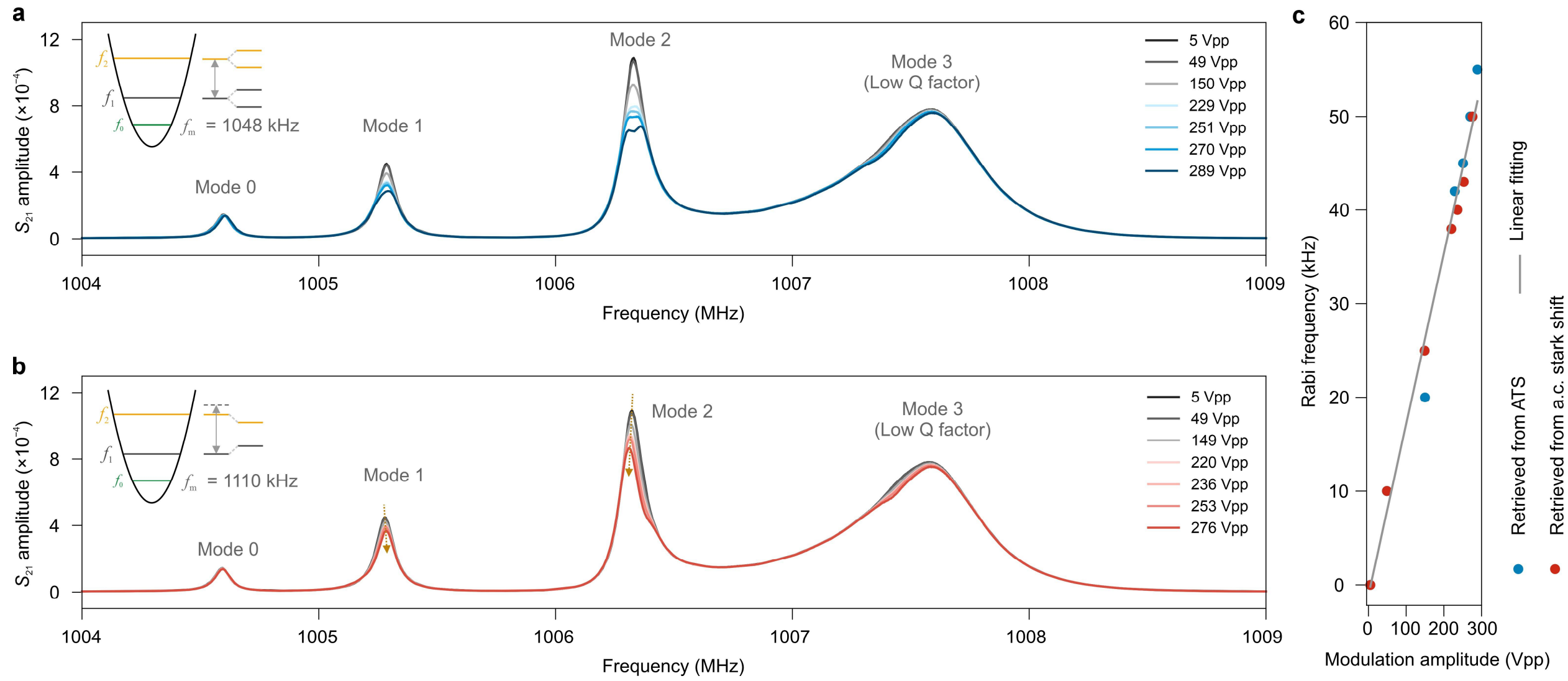

**Fig. S5. Electrical control between Mode 1 and Mode 2 of our PnC resonator.** (**a**) Measured acoustic ATS in the PnC resonator, when the frequency of the applied electrical modulation $f_m$ closely matches $f_{12}$. The acoustic ATS splitting near $f_2$ becomes larger as the modulation amplitude increases. The modulation amplitude above 1 MHz is limited by our custom high-voltage amplifier, and the splitting of Mode 1 is not experimentally observed. (**b**) Measured shift of the acoustic levels (acoustic a.c. Stark shifts), when the frequency of the electrical modulation is 62 kHz larger than $f_{12}$. For such a blue-detuned electrical modulation signal, i.e. $f_m > f_{12} > 0$, the two acoustic energy levels move towards each other. The coupling between Mode 2 and Mode 3 is a minor effect since the $Q$ factor of Mode 3 is relatively low. (**c**) Rabi frequencies extracted from acoustic ATS and ac. Stark shifts measurements show consistent and expected linear dependency on applied modulating amplitudes. Extracted parameters: $f_1$ =1,005.285 MHz, $f_2$ = 1,006.320 MHz, $Q_1$ = 10,600, $Q_2$ = 8,268, $\gamma_1 = 2\pi f_1/Q_1$, $\gamma_2 = 2\pi f_2/Q_2$, $\gamma_{1ext} = \gamma_1/26.5$, $\gamma_{2ext} = \gamma_2/17$.

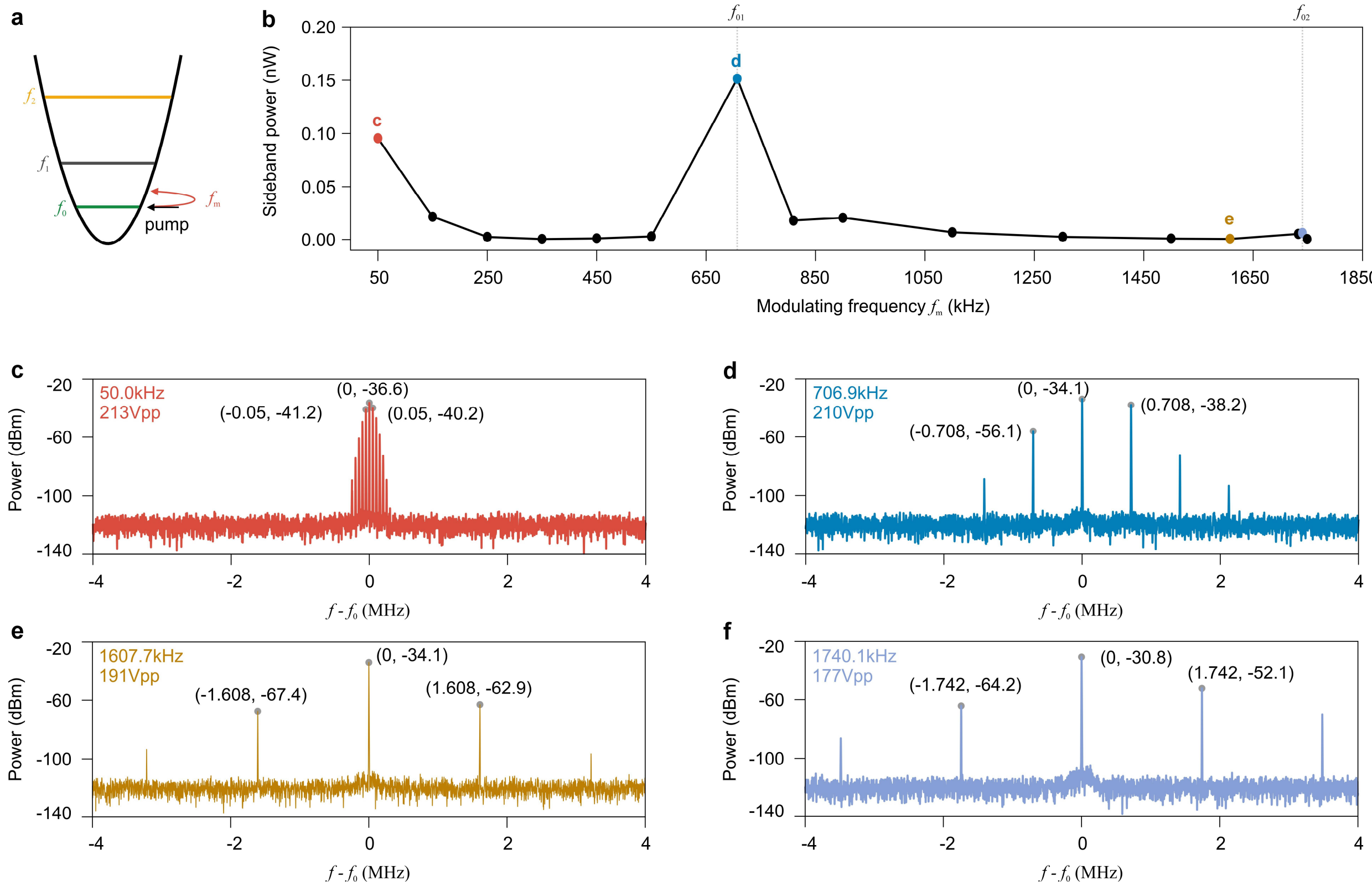

**Fig. S6. The selection rule for acoustic transitions.** (**a**) The PnC resonator is pumped with a continuous wave of $f_0$ = 1,004.460 MHz at an input power of 12 dBm. The frequency of the electrical modulation is varied from 50 kHz to 1749 kHz and its amplitude is set 197 Vpp ± 20 Vpp. (**b**) The measured power of the first sideband on the right side of the pump frequency as a function of the modulating frequency. $f_{01}$ is the frequency difference between Mode 0 and Mode 1, while $f_{02}$ is the frequency difference between Mode 0 and Mode 2. The pump signal is applied at Port 1, while the output signal is read at Port 2. (**c**) Measured power spectrum under an electrical modulation of 50 kHz at 213 Vpp. (**d**) Measured power spectrum, when the frequency of electrical modulation matches the transition frequency $f_{01}$ = 706.9 kHz. The transition between Mode 0 and Mode 1 is allowed. Single sideband modulation is achieved with an opposite sideband suppression rate of 17.9 dB. (**e**) Measured power spectrum, when the frequency of electrical modulation is 1607.7 kHz (detuned from both $f_{01}$ and $f_{02}$). The sideband power is low because of the unmatched transition frequency. (**f**) Measured power spectrum, when the frequency of electrical modulation matches $f_{02}$. The power of the first right sideband is 13.9 dB lower than that in (**d**), which indicates that the transition between Mode 0 and Mode 2 is forbidden.

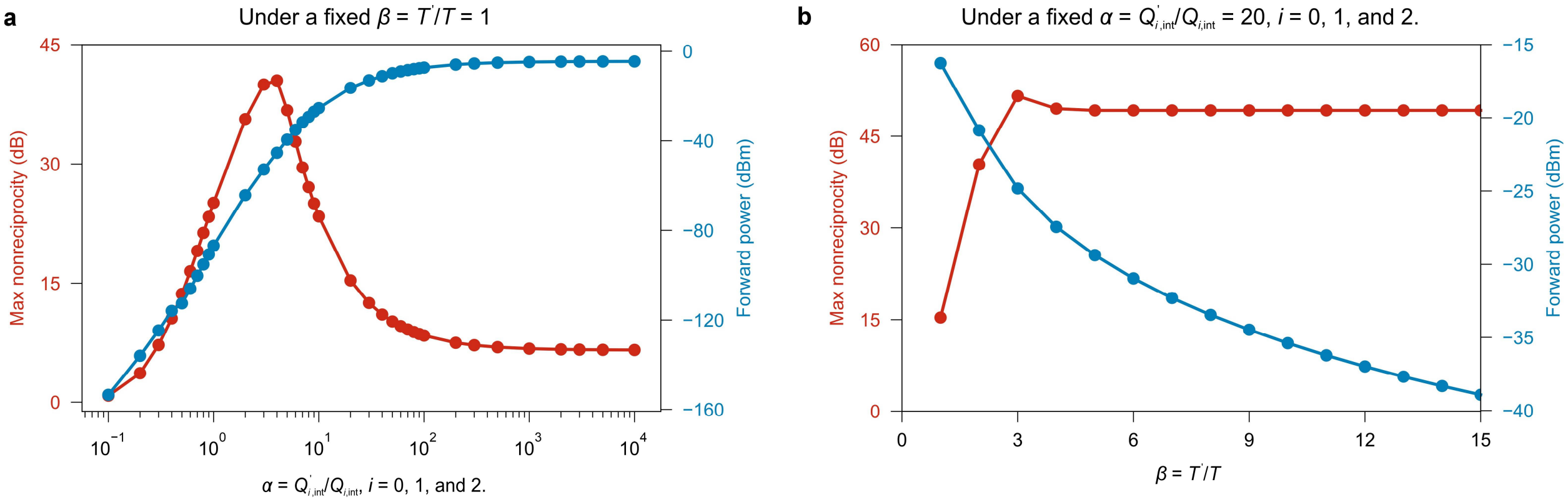

**Fig. S7. Simulated non-reciprocal frequency conversions conversion using different intrinsic *Q* factors of modes and different modulation periods *T*.** $Q_{i,\mathrm{int}}$ ($i$ = 0, 1, and 2) is the experimental intrinsic $Q$ factor of modes and $T$ is the experimental modulation period used in **Fig. 4** in the main text, while $T' = \beta T$ and $Q'_{i,\mathrm{int}} = \alpha Q_{i,\mathrm{int}}$ ($i$ = 0, 1, and 2) are the varying parameters in the numerical study. Other device parameters and modulation parameters are fixed as those in **Fig. 4** in the main text. (**a**) Maximum nonreciprocity and the corresponding forward power in the simulation as a function of the intrinsic $Q$ factors of modes $Q'_{i,\mathrm{int}}$ ($i$ = 0, 1, and 2) for $T' = T = 80$ µs. In this case, a maximum nonreciprocity of 40.5 could be realized when $Q'_{i,\mathrm{int}}$ becomes 4 times larger (i.e., $Q'_{i,\mathrm{int}} = 4Q_{i,\mathrm{int}}$, $i$ = 0, 1, and 2). Here, the loaded $Q$ is limited by the external coupling $Q$ as α becomes larger than 100. If the loaded $Q$ becomes more than 100 times larger, the max nonreciprocity would approach towards zero. (**b**) Maximum nonreciprocity and the corresponding forward power in the simulation as a function of the modulation period $T'$ for $Q'_{i,\mathrm{int}} = 20Q_{i,\mathrm{int}}$, $i$ = 0, 1, and 2. In this case, a longer modulation period $T' = 3T = 240$ µs is needed for the maximum nonreciprocity.

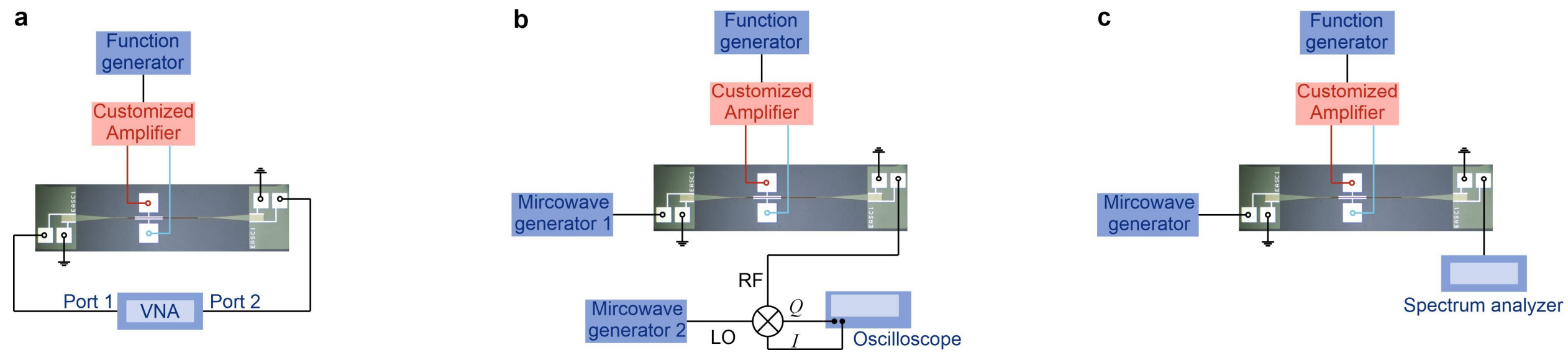

**Fig. S8. Experimental setups.** (**a**) *S*-parameters measurements. (**b)** Time domain analysis. (**c**) Spectrum analysis.

| | **Device 1[#]** | **Device 2[*]** |
|---|---|---|
| $f_0$ (MHz) | 1,004.620 | 1,003.310 |
| $f_{01}$ (MHz) | 0.686 | 0.8955 |
| $f_{12}$ (MHz) | 1.041 | 1.2465 |
| $Q_0$ | 12,020 | 10,844 |
| $Q_1$ | 10,600 | 8,296 |
| $Q_2$ | 8,268 | 6,765 |
| $\gamma_0/2\pi$ (kHz) | 83.58 | 92.52 |
| $\gamma_1/2\pi$ (kHz) | 94.84 | 121.05 |
| $\gamma_2/2\pi$ (kHz) | 121.72 | 148.63 |
| $\gamma_{0ext}$ | $\gamma_0/49.0$ | $\gamma_0/95.2$ |
| $\gamma_{1ext}$ | $\gamma_1/26.5$ | $\gamma_1/29.4$ |
| $\gamma_{2ext}$ | $\gamma_2/17.0$ | $\gamma_2/29.0$ |
| $\Omega_1/2\pi$ (maximum value achieved in experiments[+]) | 93 kHz at 336 Vpp modulation | 96 kHz at 315 Vpp modulation |
| $\Omega_2/2\pi$ (maximum value achieved in experiments[++]) | 55 kHz at 289 Vpp modulation | 47 kHz at 238 Vpp modulation |
| $C_1$ (maximum value achieved in experiments) | 4.18 | 3.29 |
| $C_2$ (maximum value achieved in experiments) | 1.05 | 0.49 |

**Table S1. Device parameters**

Devices 1 and 2 are the same design but have slightly different resonant parameters due to fabrication tolerances.

# For **Figs. 1** to **3** and **Figs**. **S2**, **S4**, **S5**, and **S6**.

* For **Fig. 4** and **Fig**. **S7**.

[+] The limitation of the maximum voltage is the air breakdown voltage of the 8 mm gap between electrodes.

[++] The limitation of the maximum voltage is the slew rate of the op-amp in our voltage amplifier.

**Movie S1. Measured out-of-plane displacement profiles of the PnC resonator.** The resonator is pumped by a continuous wave at frequency $f_0$ with power of 10 dBm and is electrically modulated by a continuous sine wave with amplitude of 210 Vpp and frequency of $f_{01}$. The top plot shows the displacement near frequency $f_0$, the second plot shows the displacement near frequency $f_0$, the third plot shows the overlay of displacements at $f_0$ and $f_1$, and the bottom plot shows the total displacement combining displacements at both frequencies. For better visualization, the displacement at frequency $f_0$ is shifted to 20 MHz, and displacement at $f_1$ is shifted to 20 MHz+$f_{01}$; this will not change the phase relationship between displacement at two frequencies. The displacement profiles are measured using our in-house vibrometer.